\documentclass[prd,nofootinbib,preprint,superscriptaddress]{revtex4}

\usepackage{graphicx}
\usepackage{amsmath}
\usepackage{amssymb}

\usepackage[hypertex]{hyperref}
\newcommand{\beq}{\begin{equation}}
\newcommand{\eeq}{\end{equation}}

\newcommand{\vect}[1]{\boldsymbol{\rm #1}}

\newcommand{\capdef}{}
\newcommand{\mycaption}[2][\capdef]{\renewcommand{\capdef}{#2}%
       \caption[#1]{{\footnotesize #2}}}
\makeatletter
\renewcommand{\fnum@table}{\textbf{\tablename~\thetable}}
\renewcommand{\fnum@figure}{\textbf{\figurename~\thefigure}}
\makeatother

\begin{document}
\pagestyle{plain}

\vspace*{1cm}
\preprint{CERN-PH-TH/2008-171}

\title{Spin-independent elastic WIMP scattering and\\ 
the DAMA annual modulation signal\vspace*{1cm}}

\author{\textbf{Malcolm Fairbairn}}
\email{malc_at_cern.ch}
\affiliation{Physics Department, Theory Division, CERN,
1211 Geneva 23, Switzerland}

\affiliation{King's College London, Strand, WC2R 2LS, UK\vspace*{5mm}}

\author{\textbf{Thomas Schwetz}\vspace*{3mm}}
\email{schwetz_at_cern.ch}
\affiliation{Physics Department, Theory Division, CERN,
1211 Geneva 23, Switzerland}


\begin{abstract}
  \vspace*{5mm} We discuss the interpretation of the annual modulation
  signal seen in the DAMA experiment in terms of spin-independent
  elastic WIMP scattering. Taking into account channeling in the
  crystal as well as the spectral signature of the modulation signal
  we find that the low-mass WIMP region consistent with DAMA data is
  confined to WIMP masses close to $m_\chi \simeq 12$~GeV, in
  disagreement with the constraints from CDMS and XENON. We conclude
  that even if channeling is taken into account this interpretation of
  the DAMA modulation signal is disfavoured. There are no overlap
  regions in the parameter space at 90\%~CL and a consistency test
  gives the probability of $1.2\times 10^{-5}$. We study the
  robustness of this result with respect to variations of the WIMP
  velocity distribution in our galaxy, by changing various parameters
  of the distribution function, and by using the results of a
  realistic $N$-body dark matter simulation. We find that only by
  making rather extreme assumptions regarding halo properties can we
  obtain agreement between DAMA and CDMS/XENON.
\end{abstract}
\maketitle


\section{Introduction}

The DAMA collaboration has collected an impressive amount of data in
their search for the scattering of weakly interacting dark matter
particles (WIMPs) off Sodium Iodine. The combined data from DAMA/NaI
(7 annual cycles) and DAMA/LIBRA (4 annual cycles) amounts to a total
exposure of 0.82~ton~yr~\cite{Bernabei:2008yi}, in a field where
exposure is measured in units of kg~days. DAMA/LIBRA has now provided
further evidence for an annual modulation of the event rate in the
energy range between 2 and 6~keVee, the claimed statistical confidence
of the positive signal being $8.2\sigma$~\cite{Bernabei:2008yi}. The
phase of the observed modulation (with maximum on day $144\pm8$) is in
striking agreement with the expectation for a modulation in a WIMP
scattering signal due to the rotation of the Earth around the Sun,
(expected maximum day 152, June 2nd), see e.g.,~\cite{Jungman:1995df}
for a review.
An interpretation of this effect in terms of spin-independent
interactions of conventional WIMPs with masses $m_\chi \gtrsim 50$~GeV
is in direct conflict with the constraints from several experiments
looking for direct WIMP detection, most notably with the data from
CDMS~\cite{Ahmed:2008eu} and XENON10~\cite{Angle:2007uj}, which
exclude the WIMP cross section consistent with the DAMA modulation for
$m_\chi\sim 50$~GeV by many orders of magnitude.  In light of this,
several alternative explanations of the DAMA annual modulation have
been proposed, for example spin-dependent
interactions~\cite{Ullio:2000bv, Savage:2004fn}, light WIMPs with
$\lesssim 10$~GeV masses~\cite{Bottino:2003iu, Bottino:2003cz,
  Gondolo:2005hh}, keV scale axion-like dark
matter~\cite{Bernabei:2005ca} (see however, \cite{Pospelov:2008jk,
  Gondolo:2008dd}), dark matter interacting only with
electrons~\cite{Bernabei:2007gr}, inelastic WIMP
scattering~\cite{TuckerSmith:2001hy, Chang:2008gd} and mirror dark
matter~\cite{Foot:2008nw}.

In this work we reconsider the possibility of spin-independent elastic
scattering of light WIMPs with $\lesssim 10$~GeV
masses~\cite{Bottino:2003iu, Bottino:2003cz, Gondolo:2005hh},
see~\cite{Bottino:2007qg, Bottino:2008mf, Petriello:2008jj,
Feng:2008dz} for recent studies. The original idea is that light dark
matter scattering on the relatively light Sodium nuclei in DAMA could
deposit enough energy in the detector to give a signal, whereas the
scattering of light halo particles off heavier nuclei, such as for
example Ge in CDMS or Xe in XENON would lead to energy depositions
below the threshold of those detectors. Recently the importance of the
so-called channeling effect~\cite{Drobyshevski:2007zj} in the crystal
structure of the experiment has also been
emphasized~\cite{Petriello:2008jj, Bottino:2008mf}.
Specific models for WIMPs with $m_\chi \sim 10$~GeV have been studied
for example in~\cite{Bottino:2002ry, Bottino:2007qg, Barger:2005hb,
Gunion:2005rw}. Here we do not discuss theoretical implications but
focus on the phenomenology of direct detection experiments in a
model-independent way by assuming that such light WIMPs can provide the
correct relic abundance while any direct collider constraints can be
evaded.

In this region of WIMP masses several
experiments~\cite{Altmann:2001ax, Akerib:2003px, Lin:2007ka,
Aalseth:2008rx} exclude WIMP--nucleon scattering cross sections in the
range $\sigma_p \gtrsim 10^{-40}$~cm$^2$. As we will see in the next
pages, once we have included channeling as well as the spectral shape
of the DAMA modulation signal, the allowed region of our interest is
obtained at much small cross sections, around $\sigma_p \sim
10^{-41}$~cm$^2$ and $m_\chi \sim 10$~GeV. In this region the most
relevant constraints come from XENON~\cite{Angle:2007uj}, the 2008
Germanium data from CDMS~\cite{Ahmed:2008eu}, and the 2005 CDMS data
on Silicon~\cite{Akerib:2005kh}. Indeed, as we will discuss, the
spectral shape of the DAMA annual modulation restricts $m_\chi$ and
$\sigma_p$ to a region excluded by these experiments.

In our study we elaborate on this result and discuss how robust it is
with respect to different assumptions about the dark matter halo of
our galaxy. The impact of non-standard halo properties on dark matter
direct detection experiments has been discussed by many authors, see
for example~\cite{Belli:2002yt, Fornengo:2003fm, Green:2002ht,
  Vergados:2007nc}.  At a qualitative level, one would expect that
smaller velocity dispersions or truncated velocity distributions would
seem to favour the dark matter interpretation of the DAMA signal, as
they could lead to more events above the low energy threshold of DAMA
but below that of other experiments.  Furthermore, anisotropies in the
velocity dispersion could amplify annual modulation signals.

The outline of our work is as follows. In Sec.~\ref{sec:analysis} we
briefly summarise the phenomenology of elastic WIMP scattering in
direct detection experiments and give some technical details on our
analysis of DAMA, CDMS and XENON data. The results for a standard dark
matter halo are presented in Sec.~\ref{sec:std-halo}.  In
Sec.~\ref{sec:nonstd-halo} we consider deviations from the standard
assumptions made about the WIMP velocity distribution: we use results
from the Via Lactea $N$-body dark matter
simulation~\cite{Diemand:2006ik}, we vary several parameters of the
Maxwellian distribution and consider asymmetric velocity profiles.
Sec.~\ref{sec:conclusions} contains our conclusions.  In
Appendix~\ref{sec:dama-mod} we comment on the DAMA fit using the
annual modulation energy spectrum, and in
Appendix~\ref{app:comparison} we briefly compare our results to the
ones from other authors.

\section{The WIMP signal in direct detection experiments}
\label{sec:analysis}

In this section we briefly summarise the phenomenology of WIMP
scattering and describe our analysis of DAMA, CDMS and XENON data.

\subsection{The event spectrum from elastic WIMP scattering}

The differential event spectrum for WIMP scattering in counts per unit
mass of a given nucleus per unit exposure time and per unit energy as
a function of the recoil energy $E_R$ is given by the expression (see e.g.,
\cite{Jungman:1995df})
\begin{equation}\label{eq:spectrum}
R(E_R) = \frac{\rho \, \sigma_p A^2 F^2(q)}{2 m_\chi \mu^2_p} \, \eta(E_R, t) \,.
\end{equation}
Here $\rho$ is the local WIMP energy density for which we adopt the
canonical value $\rho = 0.3$~GeV/cm$^3$, $\sigma_p$ is the WIMP
scattering cross section on a proton\footnote{Note that only the
product of $\rho \times \sigma_p$ is relevant for the scattering
rate. Therefore, whenever we use the symbol $\sigma_p$ the cross
section is implicitly normalised to the value of $\rho =
0.3$~GeV/cm$^3$.}, $A$ is the mass number of the target nucleus,
$\mu_p = m_\chi m_p/(m_\chi + m_p)$ is the reduced WIMP--proton mass
and we use the common Helm form factor $F(q) = 3 e^{-q^2 s^2/2}
[\sin(qr)-qr\cos(qr)] / (qr)^3$, with $s = 1$~fm, $r = \sqrt{R^2 - 5
s^2}$, $R = 1.2 A^{1/3}$~fm, $q = \sqrt{2 M E_R}$, with $M$ being the
nucleus mass. The function $\eta$ contains the integral over the WIMP
velocity distribution:
\begin{equation}\label{eq:eta}
\eta(E_R, t) = \int d\Omega_{\vect{v}} \int_{v_\mathrm{min}(E_R)}^\infty 
dv\, v \, f_\oplus(\vect{v},t) \,,
\end{equation}
where $v_\mathrm{min} = \sqrt{M E_R/2 \mu_M^2}$ is the minimum
velocity of a WIMP to produce a recoil energy $E_R$, and $v =
|\vect{v}|$. The WIMP velocity distribution in the Earth rest frame
$f_\oplus(\vect{v},t)$ is obtained from the distribution in the
galactic rest frame $f_\mathrm{gal}(\vect{v})$ by
\begin{equation}
f_\oplus(\vect{v},t) = f_\mathrm{gal}
(\vect{v} + \vect{v}_\odot +  \vect{v}_\oplus(t)) \,.
\end{equation}
In the coordinate system in which $x$ points towards the galactic
center, $y$ towards the direction of galactic rotation, and $z$
towards the galactic north pole, we use for the velocity of the Sun
$\vect{v}_\odot = (0,220,0) + (10, 13, 7)$~km/s~\cite{Gelmini:2000dm}
(including the local Keplerian velocity of 220~km/s~\cite{pdg} as well
as the Sun's peculiar velocity, see also~\cite{Green:2003yh} and
references therein). To describe the motion of the Earth around the
Sun we use the parametrisation of~\cite{Gelmini:2000dm}:
$\vect{v}_\oplus(t) = v_\oplus(\vect{e}_1 \sin\lambda - \vect{e}_2
\cos\lambda)$, with $v_\oplus = 2\pi \mathrm{A.U./yr} = 29.8$~km/s,
$\vect{e}_1 = (-0.0670, 0.4927, -0.8676)$, $\vect{e}_2 = (-0.9931,
-0.1170, 0.01032)$, and $\lambda(t) = 2\pi(t - 0.218)$.

The ``standard halo model'' assumes for the DM distribution an
isotropic isothermal sphere, which leads to a Maxwellian velocity
distribution in the galactic frame, truncated at the escape velocity
$v_\mathrm{esc}$:
\begin{equation}\label{eq:std-halo}
f_\mathrm{gal}(\vect{v}) = \left\{
\begin{array}{l@{\qquad}l}
N \, \left[\exp\left(-v^2 / \bar v^2 \right) - 
\exp\left(-v_\mathrm{esc}^2 / \bar v^2 \right)\right]
& v < v_\mathrm{esc} \\
0 & v > v_\mathrm{esc}
\end{array}\right. \,,
\end{equation}
where we adopt as default values $\bar v = 220$~km/s and
$v_\mathrm{esc} = 650$~km/s. Here and throughout the paper we use the notation $\bar v^2 = 2 (<v^2>-<v>^2)=2\sigma^2$.  In order to properly take into account
the impact of the finite escape velocity as well as allowing for
non-standard halos deviating from Eq.~\ref{eq:std-halo} we perform the
integral in Eq.~\ref{eq:eta} numerically. In Sec.~\ref{sec:std-halo}
we first consider the standard halo model, whereas in
Sec.~\ref{sec:nonstd-halo} we go beyond these default assumptions by
varying the parameters of the velocity distribution as well as
changing its shape.

\subsection{On quenching and channeling}

In the analysis of DAMA data the effects of quenching and channeling
are important~\cite{Drobyshevski:2007zj, Bernabei:2007hw}. For
quenched events the recoiling nucleus loses its energy both
electromagnetically as well as via nuclear force interactions, where
the light yield in the scintillator comes mainly from the
electromagnetic part. To take this effect into account the event
energy is measured in equivalent electron energy (in keVee), defined
by $q \times E_R$ for the total nuclear recoil energy $E_R$ in
keV. For the elements in DAMA one has $q_\mathrm{Na} = 0.3$ and
$q_\mathrm{I} = 0.09$.
However, due to the crystalline structure of the target, for certain angles and
energies of the particles no nuclear force interactions happen and the
entire energy is lost electromagnetically. Hence, for these so-called
channeled events one has $q\approx 1$, see~\cite{Drobyshevski:2007zj,
Bernabei:2007hw}. For the fraction $f$ of channeled events relevant
for DAMA we use the parameterisation
\begin{equation}
f_\mathrm{Na}(E_R)
\approx \frac{e^{-E_R/18}}{1+0.75\,E_R} 
\,,\qquad
f_\mathrm{I}(E_R) \approx
\frac{e^{-E_R/40}}{1+0.65\,E_R}
\end{equation}
for $E_R$ in keV. These expressions reproduce to good accuracy the
curves shown in figure~4 of~\cite{Bernabei:2007hw}.  Departing from
Eq.~\ref{eq:spectrum}, the predicted spectrum in DAMA (in units of
counts/kg/day/keVee) is obtained by
\begin{equation}\label{eq:spect-DAMA}
R_\mathrm{DAMA}(E) = \sum_{X =\rm Na, I} 
\frac{M_X}{M_\mathrm{Na} + M_\mathrm{I}}
\left\{[1-f_X(E/q_X)] R_X(E/q_X) + f_X(E) R_X(E) \right\} \,,
\end{equation}
where the first term in the curled bracket corresponds to quenched
events and the second to channeled (and therefore unquenched) events.

Channeling does not occur in liquid Nobel gases like in the XENON
experiment. Since no information on channeling in Germanium and
Silicon is available for us, we do not take into account channeling in
CDMS. Note, however, that CDMS requires the coincidence of signals in
phonons and ionisation and hence, since channeled events would not
give a phonon signal they would not look like a WIMP signal defined by
the coincidence. Therefore, the fraction of channeled events
corresponds effectively to an efficiency factor reducing the effective
exposure. Hence, if channeling was indeed relevant for CDMS the final
exclusion limits would be somewhat weaker. 

In conclusion, channeling is an important effect for the
interpretation of data from direct detection experiments and we stress
the need of reliable information (probably requiring dedicated
measurements) on this effect for any solid DM detector.

\subsection{Fitting DAMA/LIBRA data}

For the model-independent analysis of DAMA data the signal as a
function of energy and time is parametrised as
\begin{equation}\label{eq:DAMAsignal}
S(E,t) = S_0(E) + A(E)\cos\omega(t-t_0) \,,
\end{equation}
with $\omega = 2\pi/1$~yr, $t_0 = 152$~day. For our analysis we use
the data on the modulation amplitude $A(E)$ for the full 0.82~ton~yr
DAMA exposure\footnote{Here and in the following we use the acronym
``DAMA'' to denote the combined DAMA/NaI + DAMA/LIBRA data, except
where explicitly noted otherwise.} given in figure~9
of~\cite{Bernabei:2008yi} in 36 bins from 2 to 20~keVee. As we will
see the spectral shape of the signal is quite important for
constraining the WIMP parameters. The prediction for the modulation
amplitude in an energy bin $i$ from $E_i^-$ to $E_i^+$ is obtained
from Eq.~\ref{eq:spect-DAMA} by
\begin{equation}
A^\mathrm{pred}_i = \int dE \frac{1}{2}
[R_\mathrm{DAMA}(E, t = 152) - R_\mathrm{DAMA}(E, t = 335)]
\, \int_{E_i^-}^{E_i^+} dE' G(E, E') \,
\end{equation}
where $G(E, E')$ is a Gaussian energy resolution function
with width~\cite{Bernabei:2008yh}
\begin{equation}\label{eq:Eres}
\sigma_E^\mathrm{DAMA} / E = 0.45/\sqrt{E \,[{\rm keVee}]} + 0.0091 \,.
\end{equation}
Then we construct a $\chi^2$ function
\begin{equation}\label{eq:chisq}
\chi^2_\mathrm{DAMA}(m_\chi, \sigma_p) = \sum_{i=1}^{36} \left(
\frac{A^\mathrm{pred}_i(m_\chi, \sigma_p) - A^\mathrm{obs}_i}{\sigma_i}
\right)^2 \,,
\end{equation}
using the experimental data points $A^\mathrm{obs}_i$ and their errors
$\sigma_i$ from figure~9 of~\cite{Bernabei:2008yi}.\footnote{Fig.~10
  of \cite{Bernabei:2008yi} shows that the $A^\mathrm{obs}_i$ are
  consistent with being Gaussian distributed, justifying the $\chi^2$
  adopted in Eq.~\ref{eq:chisq}.}  We find the best fit point for
the WIMP mass and the scattering cross section by minimising
Eq.~\ref{eq:chisq} with respect to $m_\chi$ and $\sigma_p$. Allowed
regions in the $(m_\chi, \sigma_p)$ plane at a given CL are obtained
by looking for the contours $\chi^2(m_\chi, \sigma_p) =
\chi^2_\mathrm{min} + \Delta\chi^2({\rm CL})$, where
$\Delta\chi^2({\rm CL})$ is evaluated for 2 degrees of freedom (dof),
e.g., $\Delta\chi^2(90\%) = 4.6$ or $\Delta\chi^2(99.73\%) = 11.8$.

In general the constant part of the spectrum, $S_0(E)$,
will consist of a time-averaged dark matter contribution $\langle R\rangle$
plus an un-identified background $B$:
\begin{equation}
S_0(E) = \langle R(E)\rangle + B(E) \,.
\end{equation}
In a given model such as for example WIMP scattering, the annual
modulation amplitude $A(E)$ and the averaged signal $\langle
R(E)\rangle$ are not independent. Hence, for a given fit to the data
on $A(E)$, the expected constant signal $\langle R(E)\rangle$ can be
predicted by using Eq.~\ref{eq:spect-DAMA}. In order to take this
additional information into account we use the data from figure~1
of~\cite{Bernabei:2008yi}, which shows the constant signal $S_0$ in 32
energy bins from 2 to 10~keVee for the DAMA/LIBRA detectors. For each
pair of $(m_\chi, \sigma_p)$ we calculate the expected signal from
WIMP scattering $\langle R\rangle$ in each of these energy
bins. Whenever $\langle R\rangle$ exceeds the observed rate in one of
the bins that particular values of $(m_\chi, \sigma_p)$ are not
consistent with the data and have to be excluded. Note that for event
rates of order 1~count/kg/day/keVee and the DAMA/LIBRA exposure of
0.53~t~yr statistical errors are negligible for this purpose.

\subsection{Analysis of CDMS and XENON}

In our analysis we include the constraints from CDMS 2005 data using
Silicon (CDMS-Si)~\cite{Akerib:2005kh}, which, despite the relatively
low exposure of 12~kg~day, provides good sensitivity to the low-mass
WIMP region because of the light mass of the target nucleus
($M_\mathrm{Si} \simeq 26$~GeV) and the low analysis threshold of
7~keV. Furthermore, we include CDMS 2008 data on Germanium
(CDMS-Ge)~\cite{Ahmed:2008eu} with a threshold of 10~keV. We use the
energy dependent efficiency from Fig.~2 of \cite{Ahmed:2008eu} which
reduces the total exposure of 398.7~kg~day to an effective exposure of
about 121.3~kg~day. For both, CDMS-Si and CDMS-Ge, no event has been
observed. We calculate the expected number of events $N^\mathrm{pred}
$ as a function of $m_\chi$ and $\sigma_p$ by integrating
Eq.~\ref{eq:spectrum} over the relevant energy range and scaling with
the exposure. A $\chi^2$ is constructed using the common expression
for Poisson distributed data~\cite{pdg}, which for zero observed
events simply becomes
\begin{equation}\label{eq:chisq-cdms}
\chi^2_\mathrm{CDMS} = 2 N^\mathrm{pred} \,. 
\end{equation}
Exclusion contours are defined by the standard $\Delta\chi^2$ cuts for
2~dof with respect to the minimum, which of course occurs for
$N^\mathrm{pred} = 0$. Conceptually this prescription differs from the
usual way to set a limit on $\sigma_p$ for fixed $m_\chi$ by requiring
$N^\mathrm{pred} < 2.3$ for a 90\%~CL limit. However, by accident,
since $\Delta\chi^2(90\%) = 4.6$ for 2~dof, in practice our $\chi^2$
definition in Eq.~\ref{eq:chisq-cdms} leads to the same exclusion
contour as the more conventional method of setting a limit on
$\sigma_p$.

For the analysis of data from the XENON10
experiment~\cite{Angle:2007uj} (XENON for brevity) we proceed in the
following way. Using the 7 bins in nuclear recoil energy from 4.5 to
26.9~keV of table~1 of~\cite{Angle:2007uj} the predicted number of
events in bin $i$, $N^\mathrm{pred}_i(m_\chi, \sigma_p)$ can be
calculated by integrating Eq.~\ref{eq:spectrum} and scaling with the
exposure 316~kg~day as well as the bin dependent efficiencies
$\epsilon_{c}$ and $A_{nr}$ given in table~1 of~\cite{Angle:2007uj}.
After the publication of Ref.~\cite{Angle:2007uj} the so called
parameter $\mathcal{L}_\mathrm{eff}$ relevant for the nuclear recoil
energy scale in XENON was remeasured~\cite{Sorensen:2008ec}. Whereas
in~\cite{Angle:2007uj} a constant value $\mathcal{L}_\mathrm{eff} =
0.19$ was used, Fig.~3 of~\cite{Sorensen:2008ec} shows an energy
dependent deviation of $\mathcal{L}_\mathrm{eff}$ from that value. We
use the information from this figure to correct nuclear recoil
energies in XENON. This leads to a somewhat higher energy threshold of
about 5.5~keV (instead of 4.5), which shifts the bound on DM
parameters to slightly higher values of $m_\chi$.

XENON observes 10 candidate events whose
recoil energies can be inferred from figure~3
of~\cite{Angle:2007uj}. They are distributed over the 7 bins as $(D_i)
= (1, 0, 0, 0, 3, 2, 4)$, with an expected background $(B_i) =
(0.2, 0.3, 0.2, 0.8, 1.4, 1.4, 2.7)$. We use the $\chi^2$ for
Poisson distributed data~\cite{pdg}:
\begin{equation}\label{eq:chisq_xenon}
\chi^2_\mathrm{XENON} = 2 \sum_{i=1}^7 \left[ 
N^\mathrm{pred}_i + B_i - D_i + 
D_i \log\left(\frac{D_i}{N^\mathrm{pred}_i + B_i}\right) \right] \,, 
\end{equation}
where the second term in the square bracket is zero if $D_i = 0$.
Again we define the exclusion curve in the $(m_\chi, \sigma_p)$ plane
by $\Delta\chi^2({\rm CL})$ contours for 2~dof with respect to the
minimum.

In both cases, CDMS and XENON we include an energy resolution of 
$20\% / \sqrt{E_R \, [{\rm keV}]}$, and an uncertainty in the energy
scale of 10\%, which is added to the $\chi^2$ definitions
Eqs.~\ref{eq:chisq-cdms} and \ref{eq:chisq_xenon} with the help of
nuisance parameters. Note that CDMS and XENON report their results
directly in terms of the recoil energy already corrected by the
quenching factor. In contrast to DAMA, here this is possible
because only a single element is used as target.

\section{Standard halo results}
\label{sec:std-halo}

\begin{figure}
\centering \includegraphics[width=0.7\textwidth]{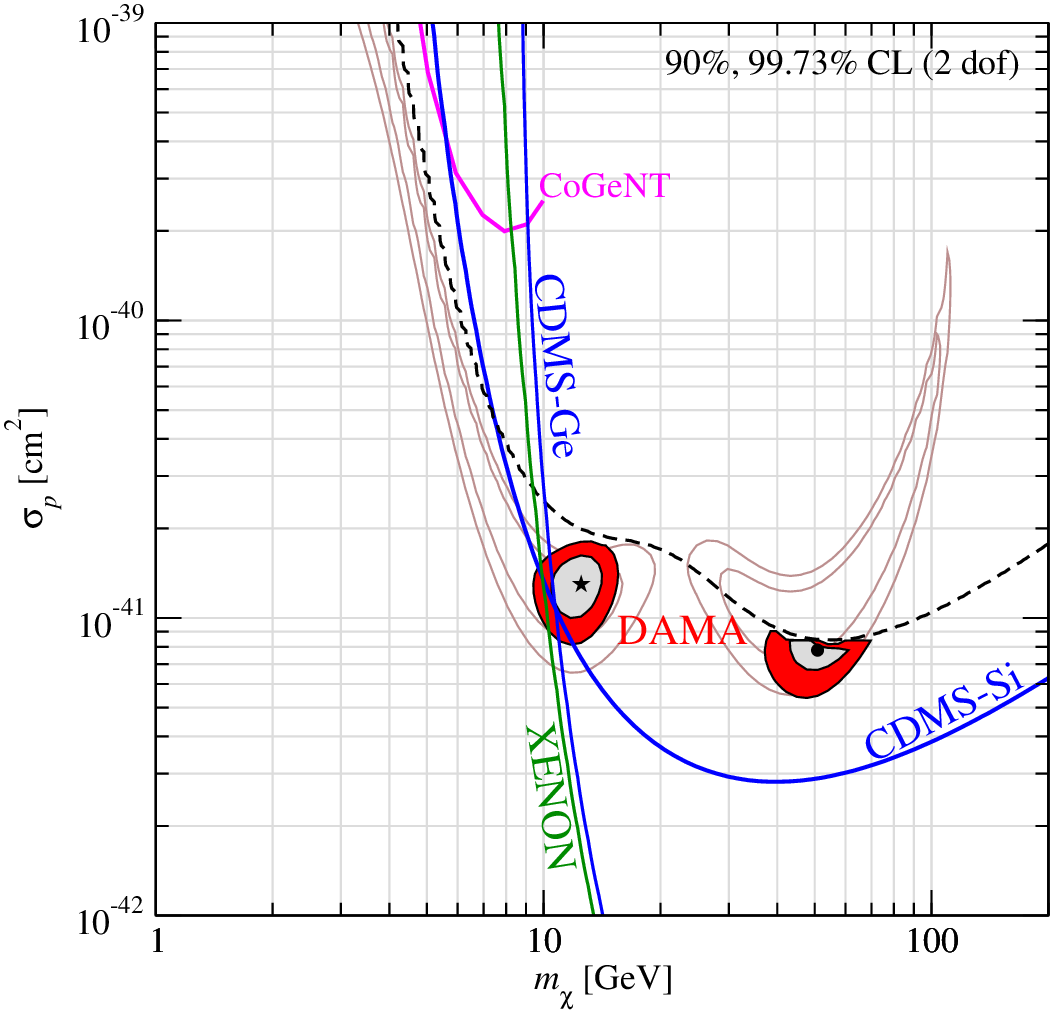}
  \mycaption{\label{fig:regions} Allowed regions at 90\% and
  99.73\%~CL for WIMP mass and scattering cross section on nucleon for
  DAMA, and exclusion contours for CDMS-Si, CDMS-Ge and XENON at
  90\%~CL. We also display the limit from CoGeNT extracted from figure~2
  of~\cite{Aalseth:2008rx}.  The global best fit for DAMA is marked
  with a star, the allowed region around $m_\chi \simeq 50$~GeV is
  defined with respect to the local minimum, which is marked with a
  dot. For DAMA we show the regions obtained from using only the
  modulation amplitude for 2--6~keVee (gray curves) and from using the
  spectral shape of the modulation signal (shaded regions). For
  parameters above the dashed curve the predicted number of events in
  DAMA/LIBRA is larger than the observed number of events.}
\end{figure}

Figure~\ref{fig:regions} summarises our results assuming standard halo
properties, showing the allowed region from DAMA together with the
constraints from CDMS-Si, CDMS-Ge and XENON. First we discuss the fit
to DAMA data alone (without constraints from CDMS and XENON).  We find
two islands in the $(m_\chi, \sigma_p)$ plane where DAMA can be
accommodated. The best fit point is obtained at
\begin{equation}\label{eq:dama-bfp}
m_\chi = 12\,{\rm GeV}\,,\quad
\sigma_p = 1.3\times 10^{-41}\,{\rm cm}^2 \,,\qquad
\chi^2_\mathrm{DAMA,min} = 36.8/34\,{\rm dof} \,,
\end{equation}
with an excellent goodness of fit of 34\%. There is also a local
minimum at $m_\chi = 51$~GeV with $\chi^2_\mathrm{local} = 47.9$. This
solution is disfavoured with respect to the best fit point at about
$3\sigma$ for 2~dof ($\Delta\chi^2 = 11.1$). The allowed regions
around $m_\chi \simeq 50$~GeV shown in figure~\ref{fig:regions} are
defined with respect to the local minimum. The low and high WIMP-mass
solutions correspond to channeled and quenched scatterings on Iodine,
respectively. In contrast to the situation when all events are assumed
to be quenched~\cite{Gondolo:2005hh}, it turns out that scattering on
Sodium is not relevant once channeling of Iodine events takes
place. The reason is that quenched events on Sodium require a similar
WIMP mass as channeled events on Iodine (i.e., $m_\chi \simeq 10$~GeV)
but a much larger cross section $\sigma_p$ (due to the $A^2$
dependence of the total cross section on the nucleus), and therefore,
are highly suppressed once channeled scattering on Iodine takes
place. In principle there would be also a solution from channeled
events on Na, around $m_\chi \simeq 5$~GeV. However, it turns out that
in this case the un-channeled events on Na still contribute to the signal,
and indeed prevent fitting the data with the channeled Na events. Note
that the solution around $m_\chi \simeq 50$~GeV is excluded by some
orders of magnitude by XENON and CDMS-Ge, and therefore we focus in
the following on the low-mass region $m_\chi \simeq 10$~GeV.

The gray contours in figure~\ref{fig:regions} correspond to an
alternative method of fitting DAMA. Instead of using the detailed
spectral information of the annual modulation, we fit the time
dependence of their signal integrated over energy. In figure~6
of~\cite{Bernabei:2008yi} data on the residual rate $(S(t) - S_0)$
(c.f., Eq.~\ref{eq:DAMAsignal}) is given in 7~time bins of one single
annual cycle. For the gray contours in figure~\ref{fig:regions} we use
these data for the energy intervals 2 to 6~keVee and 6 to 14~keVee,
where in the latter interval data are consistent with no annual
variation. These results are very similar to the ones
of~\cite{Gondolo:2005hh} (if channeling is neglected, not shown in the
figure) and~\cite{Petriello:2008jj} (including channeling), where only
two data points for the modulation amplitude below and above 6~keVee
have been used.

\begin{figure}
\centering 
\includegraphics[height=0.45\textwidth]{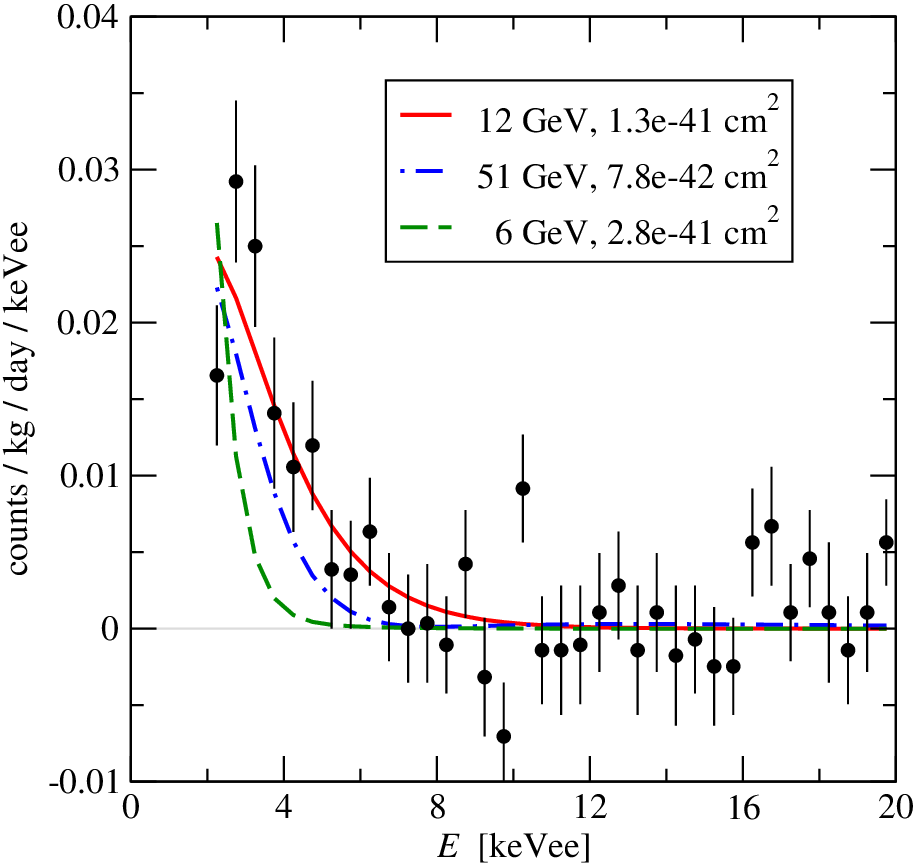} \quad
\includegraphics[height=0.45\textwidth]{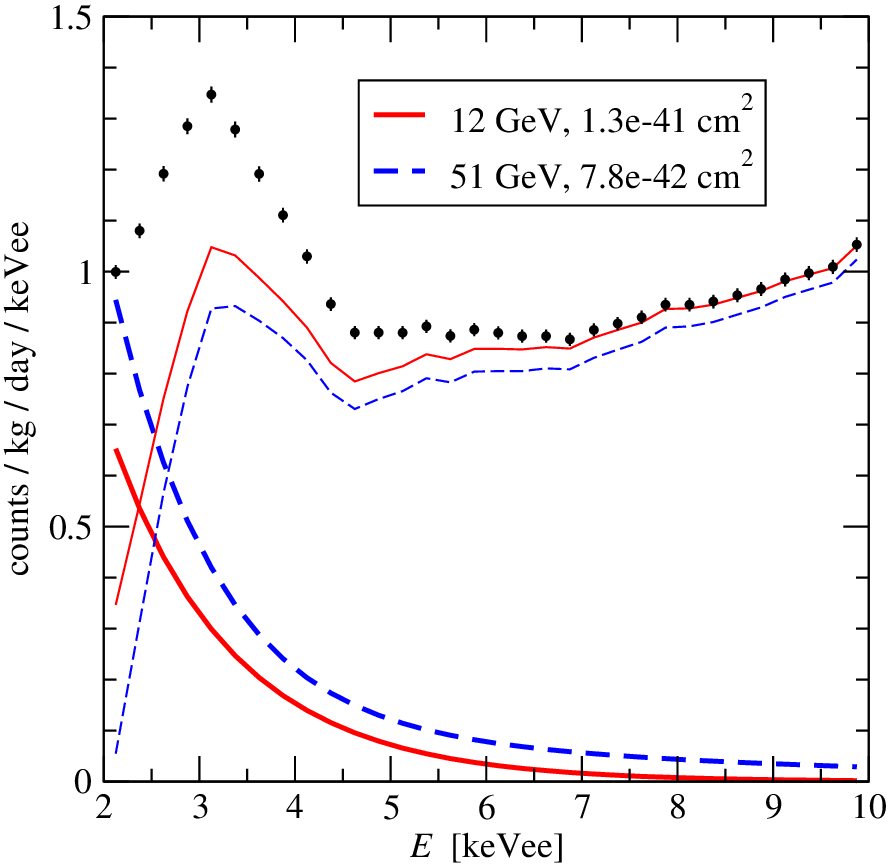}
  \mycaption{\label{fig:spectrum} Left: Energy distribution of the
  annual modulation amplitude from DAMA/NaI and DAMA/LIBRA data
  extracted from figure~9 of~\cite{Bernabei:2008yi} (points with error
  bars), together with the prediction for three examples of WIMP
  masses and scattering cross sections (curves). Right: Energy
  distribution of the time averaged rate observed in DAMA/LIBRA
  extracted from figure~1 of~\cite{Bernabei:2008yi} (points), together
  with the prediction for two examples of WIMP masses and scattering
  cross sections (thick curves) as well as the corresponding
  un-identified background (thin curves). The data are corrected for
  the energy dependent efficiency.}
\end{figure}

We observe from figure~\ref{fig:regions} that the two methods of
analysing DAMA data are consistent with each other (as it should
be), but also that using the spectral information gives significantly
stronger constraints on the allowed region. This is illustrated in
figure~\ref{fig:spectrum} (left), showing the 36 data points on the
modulation amplitude $A_i$ used in our default analysis. 
While the prediction from the best fit point of Eq.~\ref{eq:dama-bfp}
nicely follows the data (solid curve), moving to smaller WIMP masses
leads to a modulation signal more peaked at the lowest
energies. Therefore, although it is still possible to obtain the
integrated signal in the interval from 2 to 6~keVee, the spectral
shape is clearly inconsistent with data, as illustrated for $m_\chi =
6$~GeV by the dashed curve.\footnote{The value for the cross section
$\sigma_p = 2.8\times 10^{-41}$~cm$^2$ formally gives the best fit to
the data shown in figure~\ref{fig:spectrum} (left) for $m_\chi =
6$~GeV. However, the value required to obtain the integrated
modulation amplitude for this $m_\chi$ is about a factor 2 larger,
$\sigma_p = 6\times 10^{-41}$~cm$^2$, as can be seen from the gray
contours shown in figure~\ref{fig:regions}.}

Finally we mention the implication of the data on the time averaged
rate observed in DAMA. Parameter values above the dashed curve in
figure~\ref{fig:regions} are excluded because they would lead to a
higher event rate than observed. This leads to additional constraints
for the high-mass solution. In figure~\ref{fig:spectrum} (right) we show
the observed rate together with the predictions for the two local
minima. Note that for the DAMA/LIBRA exposure of 0.53~t~yr statistical
errors are not visible at the scale of the plot. Clearly, solutions
predicting a relatively large rate require that the un-identified
background drops rapidly in order to give space for the WIMP
signal. In particular, the solution at $m_\chi = 51$~GeV requires that
the background drops to zero in the first energy bin.  Although this
cannot be excluded a priori, at least such a background shape seems
somewhat unlikely. The issue is less severe for the best fit point at
$m_\chi = 12$~GeV, since the ratio of modulation amplitude to average
rate increases for decreasing WIMP mass. However, any point close to
the dashed line in figure~\ref{fig:regions} is affected by this problem.

\bigskip 

From figure~\ref{fig:regions} we find that the parameters allowed by
DAMA data at 90\%~CL are excluded by the 90\%~CL limits of CDMS-Si,
CDMS-Ge, and XENON. If all data are combined by adding the individual
$\chi^2$ functions,
\begin{equation}
\chi^2_\mathrm{global} = \chi^2_\mathrm{DAMA} + \chi^2_\mathrm{CDMS-Ge}
+ \chi^2_\mathrm{CDMS-Si} + \chi^2_\mathrm{XENON} \,,
\end{equation}
we find the minimum at $m_\chi = 9.5$~GeV and $\sigma_p = 1.2\times
10^{-41}$~cm$^2$ with $\chi^2_\mathrm{global,min} = 59.3/(45-2)\,\rm
dof$, which corresponds to a 5\% goodness of fit.

Let us note that the goodness of fit test based on
$\chi^2_\mathrm{min}/$dof often is not very sensitive to tensions in
the fit, especially in case of a large number of data points. This can
happen if there are many data points which actually are not sensitive
to the relevant parameters, and hence, allow to ``hide'' the problem
in the fit, see for example the discussion in~\cite{Maltoni:2003cu}.
Because of this the goodness of fit depends also on the way of binning
the data. To circumvent this weakness of the standard goodness of fit
test the so-called Parameter Goodness of fit (PG) can be
used~\cite{Maltoni:2003cu}. Whereas the standard test measures the
probability that all individual data points are fitted by an
hypothesis, the PG tests the consistency of different data sets under
an hypotesis. It is based on the $\chi^2$ function
\begin{equation} \label{eq:PG}
    \chi^2_\text{PG} =
    \chi^2_\text{global,min} - \sum_i \chi^2_{i,\text{min}} \,,
\end{equation}
where $\chi^2_\text{global,min}$ is the $\chi^2$ minimum of all data
sets combined and $\chi^2_{i,\text{min}}$ is the minimum of the data
set $i$. This $\chi^2$ function measures the ``price'' one has to pay
by the combination of the data sets compared to fitting them
independently. It follows a standard $\chi^2$ distribution and should
be evaluated for the number of dof corresponding to the number of
parameters in common to the data sets, see~\cite{Maltoni:2003cu} for a
precise definition.

To apply this method we consider the two data sets DAMA versus all the
other data showing no evidence. Hence, we combine CDMS-Ge, CDMS-Si,
and XENON into one data set which we denote by NEV. Then we find
$\chi^2_\text{PG} = 22.6$. Evaluating this for 2~dof (corresponding to
the two parameters $m_\chi$ and $\sigma_p$ in common to both data
sets) one finds that DAMA and NEV data are consistent only at a
probability of $1.2\times 10^{-5}$. This corresponds roughly to the
probability of a $2.9\sigma$ fluctuation in both data sets at the same
time.\footnote{\label{ft:gof} Note that the standard
  $\chi^2_\mathrm{min}$/dof probability of 5\% tests the fit of all
  individual data points at the best fit solution, whereas the PG of
  $1.2\times 10^{-5}$ reflects the compatibility of the DAMA and NEV
  data sets. These are different questions and therefore the two
  probabilities are consistent with each other. Our DAMA analysis is
  based on 36 data points on the modulation amplitude, where most of
  them (above 6~keVee) are always fitted perfectly, irrespectively of
  the disagreement with CDMS/XENON. This is an example of the above
  mentioned ``dilution'' of the standard goodness of fit test.} We
conclude that the explanation of DAMA results in terms of
spin-independent elastic scattering of WIMPs with standard halo
properties is strongly disfavoured by XENON and CDMS data. Next we
investigate the stability of this result with respect to modifications
of the velocity distribution of the WIMPs in the halo of our galaxy.

\section{Non-standard halos}
\label{sec:nonstd-halo}

The precise limits on the cross section and mass of a dark matter
candidate which are obtained from a particular
observation/non-observation of a signal in a direct detection experiment
depend upon the velocity dispersion of dark matter around the detector, and consequently ultimately upon astrophysical assumptions.  The same set of astrophysical assumptions are
normally made by different experiments so that their results can be
compared with each other, the first of which being that the dark matter halo of the
galaxy is an isothermal sphere which means a spherically symmetric density distribution of the form $\rho\propto r^{-2}$.  For such a density distribution the Keplerian velocity is independent of radius and the value of this velocity normally assumed for dark matter studies is
a radius independent Keplerian velocity of 220 kms$^{-1}$.  It is also
assumed that the velocity dispersion of the dark matter profile is everywhere
isotropic and Gaussian, the width of the Gaussian distribution
corresponding to the Keplerian velocity of the profile.  It turns out
that we do not actually expect any of these assumptions to hold true
for a realistic dark matter halo.

Over the past decade, $N$-body simulations of increasingly large
numbers of dark matter particles have allowed us to obtain more
information about the kind of dark matter halos that one would expect
to form in an expanding universe (see e.g.~\cite{Springel:2005nw}).
These simulations show that one might expect a dark matter density
that decreases more steeply with radius at large radii rather than
have the same power law at all radii as in an isothermal profile
\cite{Navarro:1995iw}.  Furthermore, the orbits taken by dark matter
particles in a realistic simulation are usually rather radial, resulting in an anisotropic
velocity dispersion \cite{Gustafsson:2006gr}.  

Note that one can also obtain an anisotropic velocity dispersion in a halo where the density distribution is not spherically symmetric, but rather triaxial as in \cite{Green:2002ht}.  In this work however, we only consider dark matter halos where the density distribution is spherically symmetric although the velocity dispersion does not have to be.
Also, in a non-extensive ideal gas where there is a long
range attractive force between the particles such as we have here in
the form of gravity, one generically expects deviations from Gaussian
velocity distribution \cite{Vergados:2007nc}.  Furthermore, if the dark matter is still not completely virialised but is still coming into equilibrium,  there will be a superposition of multiple dark matter populations at any given place in the halo.  This effect would also lead to deviations from a Gaussian distribution of velocities.

\begin{figure}
\centering 
\includegraphics[height=0.345\textwidth]{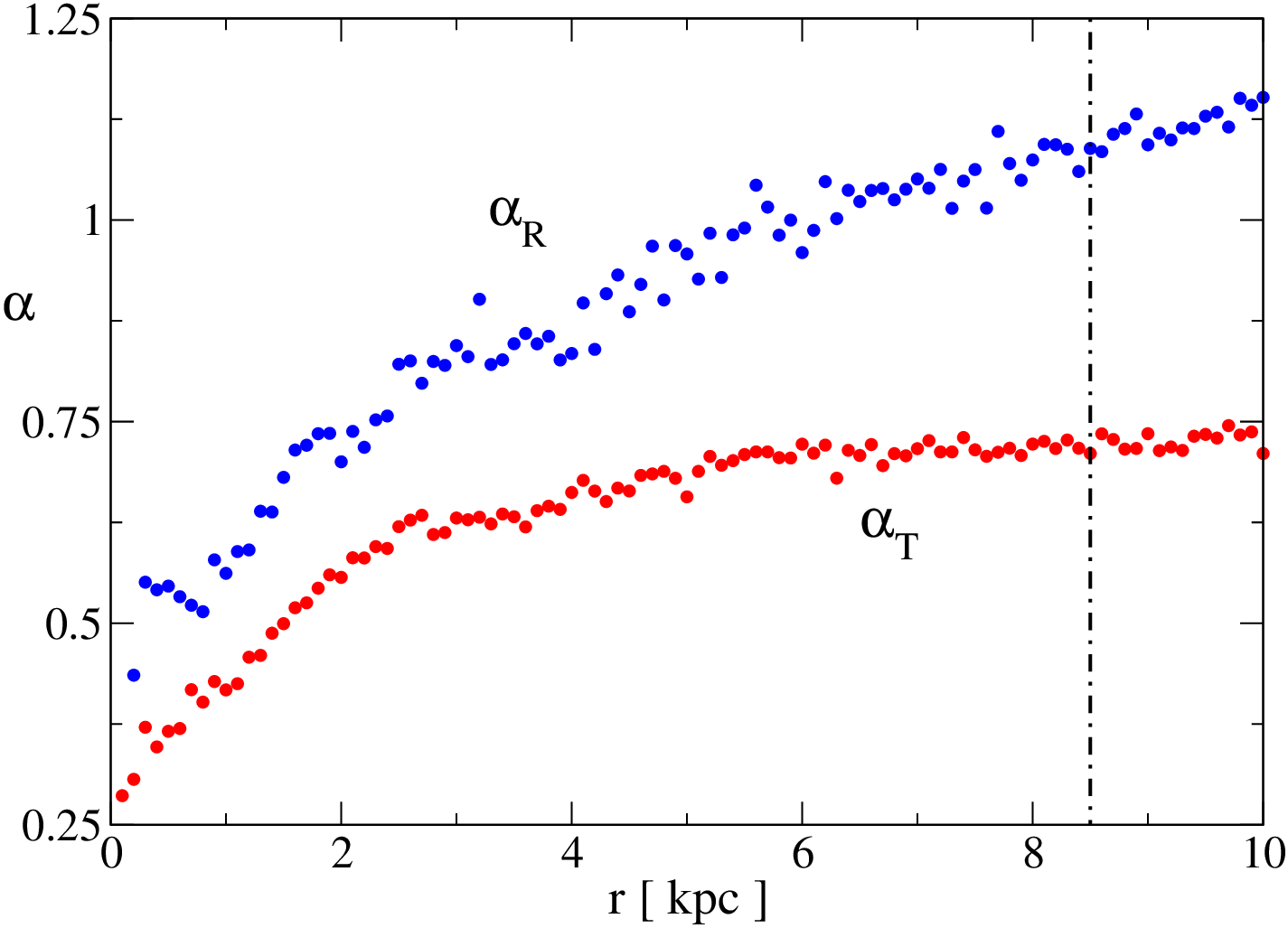} \quad
\includegraphics[height=0.345\textwidth]{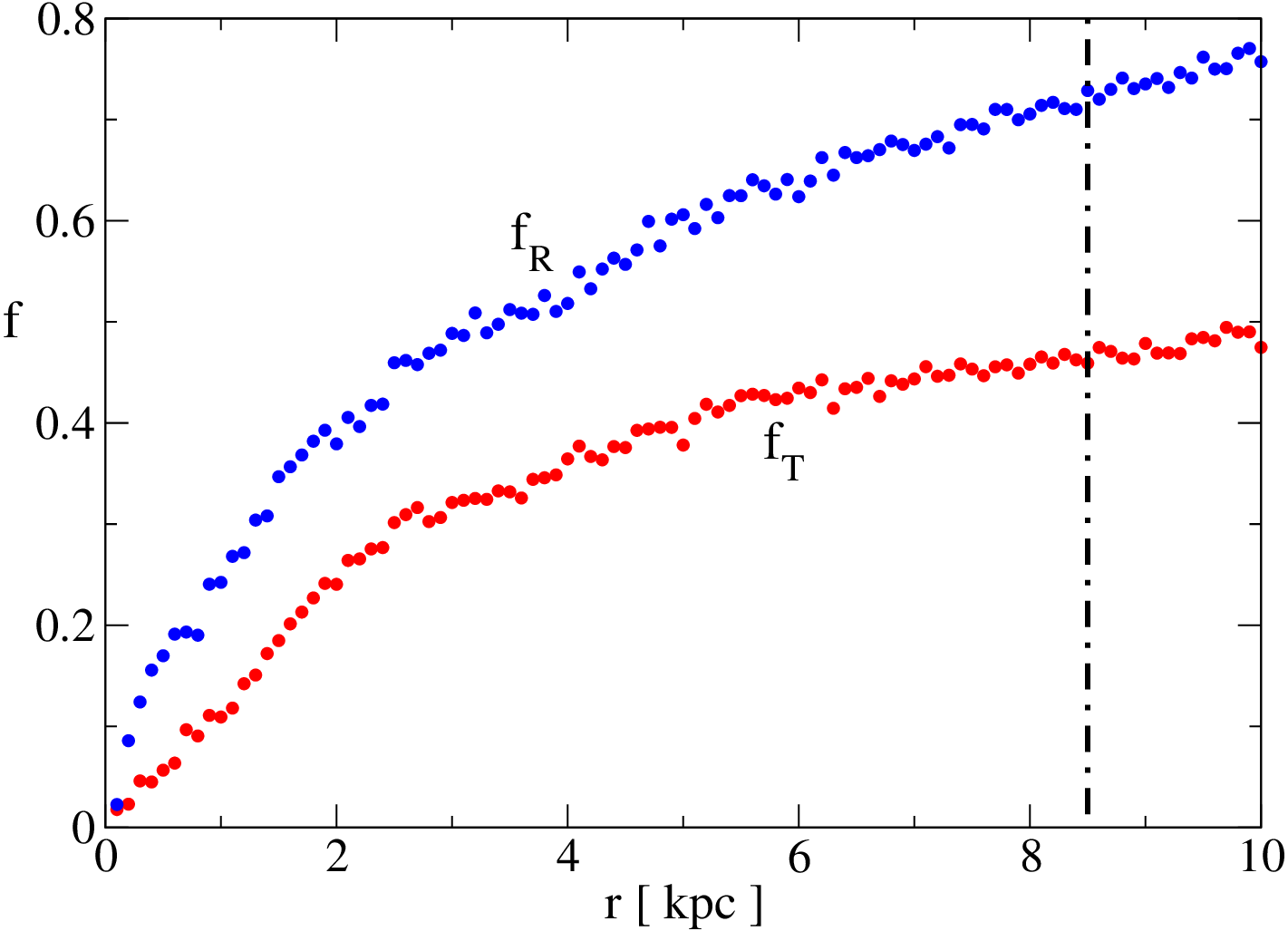}
  \mycaption{\label{fig:vl} The parameters $\alpha_i$ and $f_i$
  explained in the text fitted to the radial and tangential velocity
  dispersions of dark matter at different radii from the centre of the
  Via Lactea simulation.  The vertical lines indicate the position of
  the Sun at $r=8.5$~kpc. The velocity dispersions are clearly
  non-Gaussian as one approaches the centre of the galaxy.}
\end{figure}

In 2006, the results from a Milky Way size dark matter halo simulation
called Via Lactea containing 234 million particles were published
\cite{Diemand:2006ik}.  We have looked at this data
to see how much the dark matter distribution experienced by an
observer at the Solar radius within this simulation would vary from
the normal assumptions stated above for observers on Earth.
For each particle in the simulation there is a position vector $x_i$
and velocity vector $v_i$, plus the local gravitational potential per
unit mass in units of velocity squared $U(x_i)$.  We express the
velocities in terms of the square root of their local potential
$\tilde{v}_i=v_i/\sqrt{-U(x_i)}$.
Next we work out the angle between
the radial direction and the overall velocity vector.  We then use
this to decompose the velocity into a radial part and a part
perpendicular to that which we call tangential.  For each radius we
bin the tangential and the radial velocities obtaining two
distributions.  We fit the one dimensional radial distribution using
the following expression which we find to be a better fit than the
Tsallis distributions (designed to fit non-extensive or multiple temperature distributions) used in \cite{Vergados:2007nc}
\begin{equation}
\frac{1}{N_R}\exp\left[-\left(\frac{\tilde{v}_R^2}{f_R^2}\right)^{\alpha_R}\right] \,.
\end{equation}
Because we have rescaled the velocities with respect to $\sqrt{-U(x_i)}$, $f_R$ and $\alpha_R$ are dimensionless constants of order one.  The naive assumption is that the width of the velocity distribution at a given radius is simply the Keplerian velocity at that radius.  While information about the mass distribution of dark matter is required to go from the potential to the Keplerian velocity, The parameter $f_R$ is an indication of how badly this assumption is broken.  $\alpha$ encodes the deviation from
Gaussianity ($\alpha=1$ corresponding to a Gaussian).  For the one
dimensional case, the normalisation is analytic, $N_R = 2f_R
\Gamma(1+1/2\alpha_R)$.  We perform the same fitting procedure for the
tangential velocity, fitting
\begin{equation}
\frac{2\pi v_T}{N_T}
\exp\left[{-\left(\frac{\tilde{v}_T^2}{f_T^2}\right)^{\alpha_T}}\right]
\end{equation}
and while we are not aware of an analytic expression for $N_T$ it is
trivial to obtain it numerically. Note, in terms of the two dimensions
perpendicular to the radial direction $R$,
$\tilde{v}_T^2=\tilde{v}_\theta^2+\tilde{v}_\phi^2$.

\begin{figure}
\centering 
\includegraphics[height=0.4\textwidth]{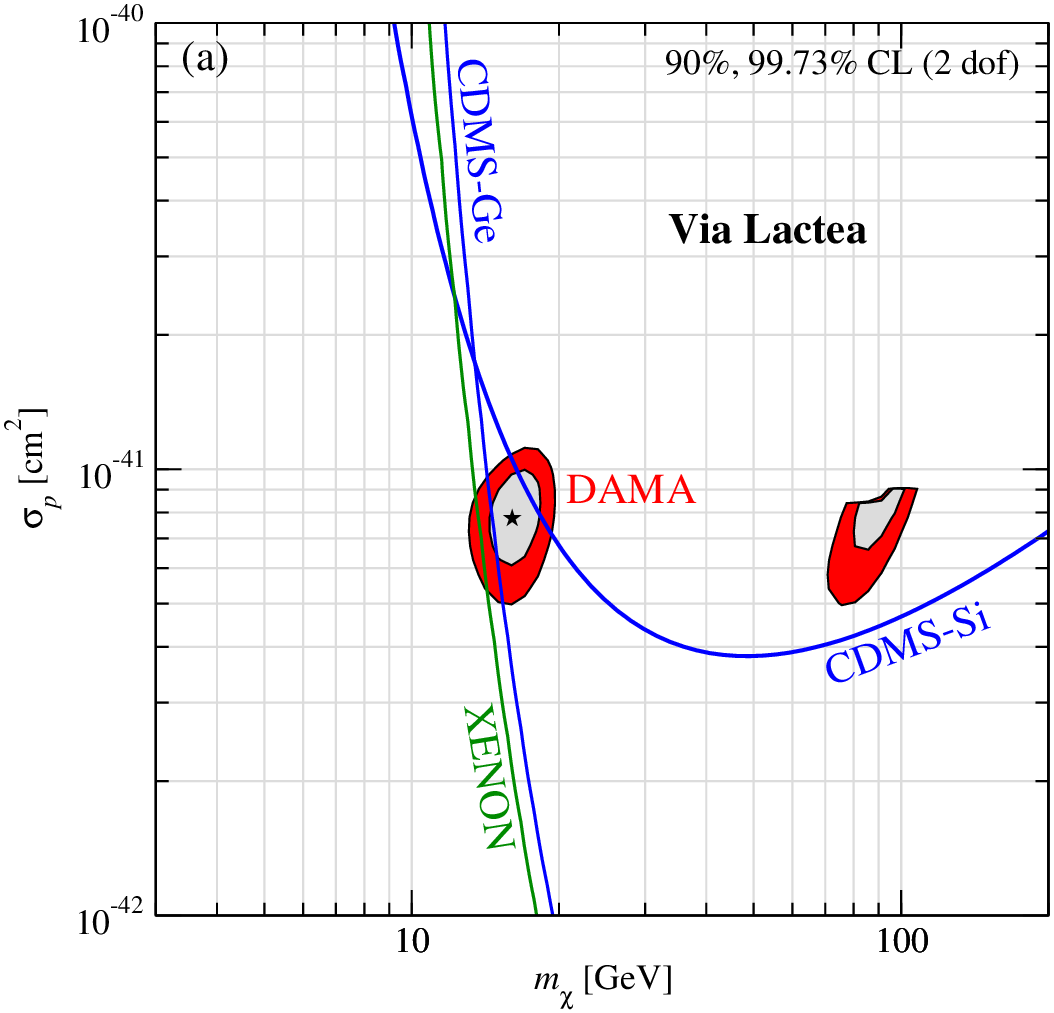} \qquad
\includegraphics[height=0.4\textwidth]{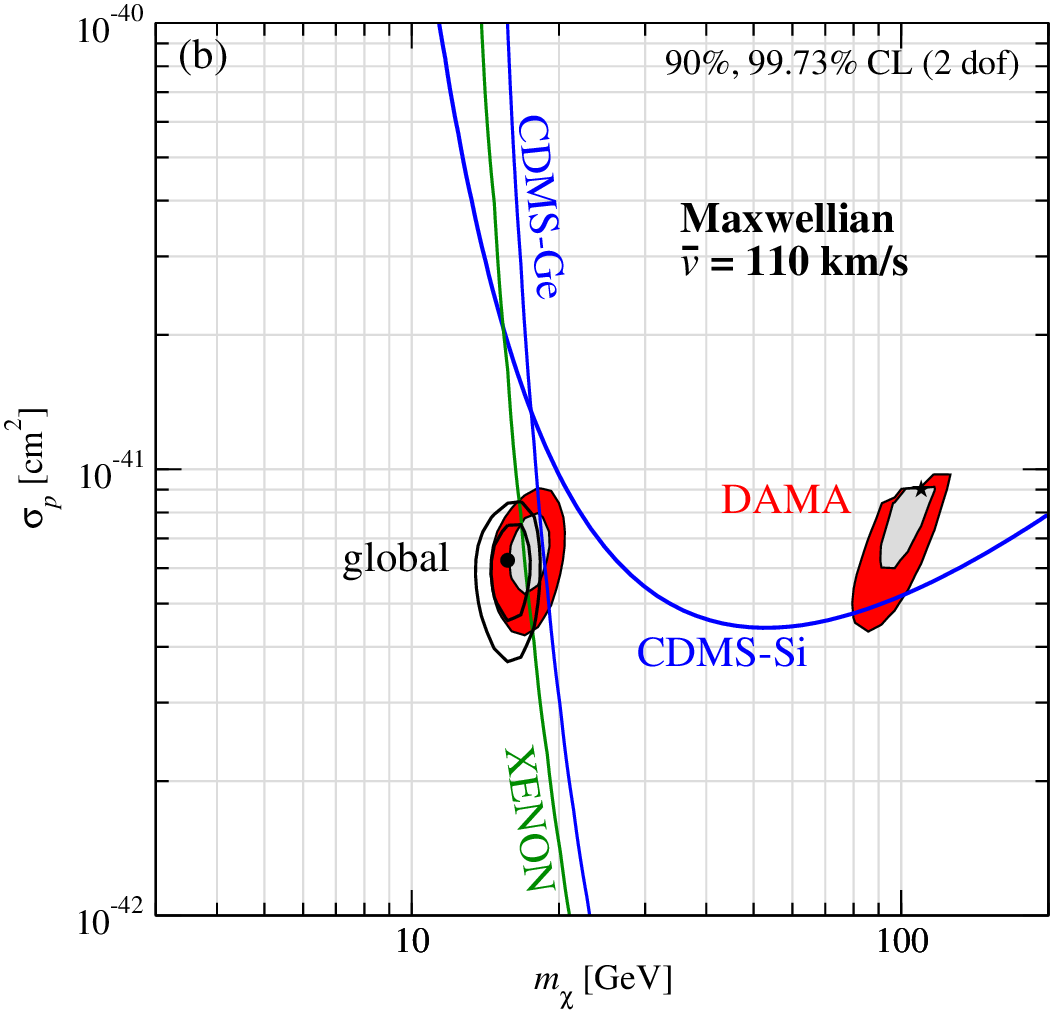} \\
\includegraphics[height=0.4\textwidth]{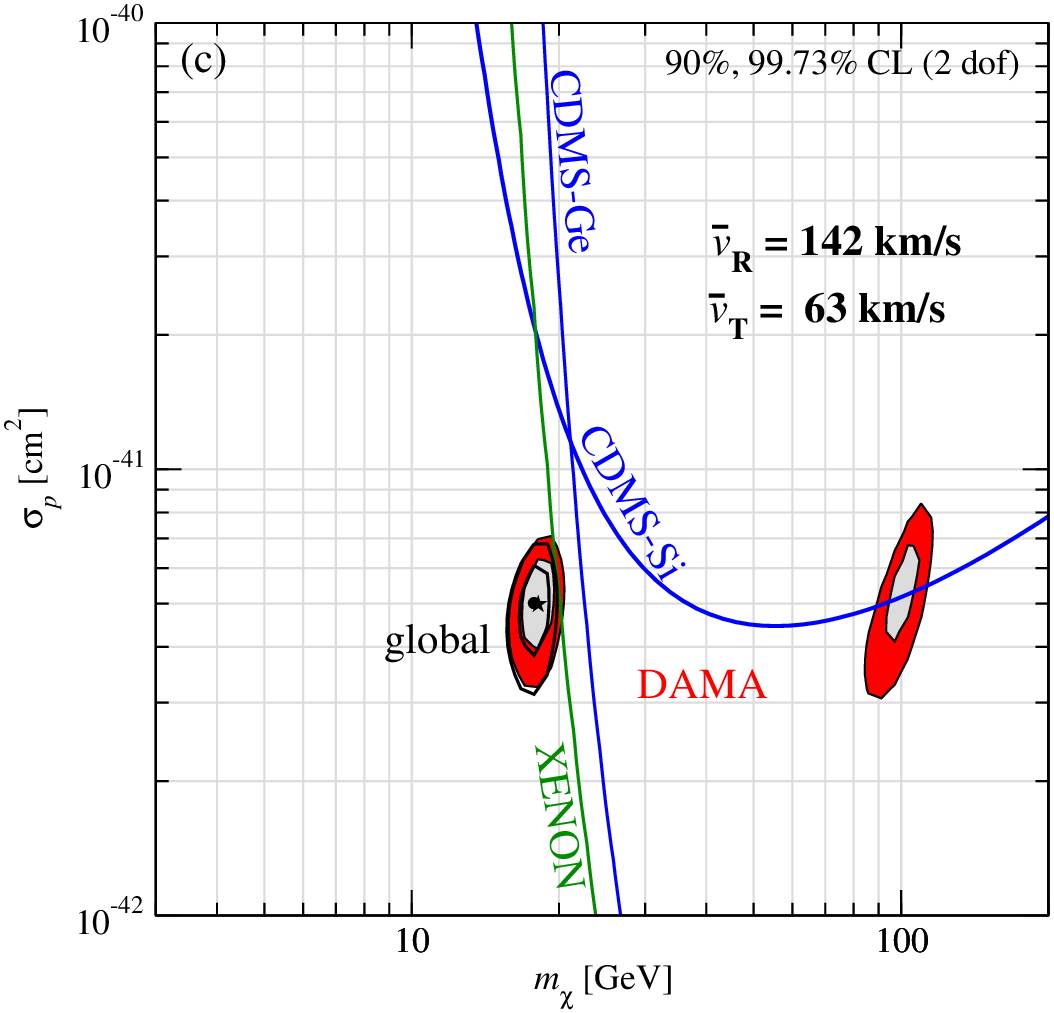} \qquad
\includegraphics[height=0.4\textwidth]{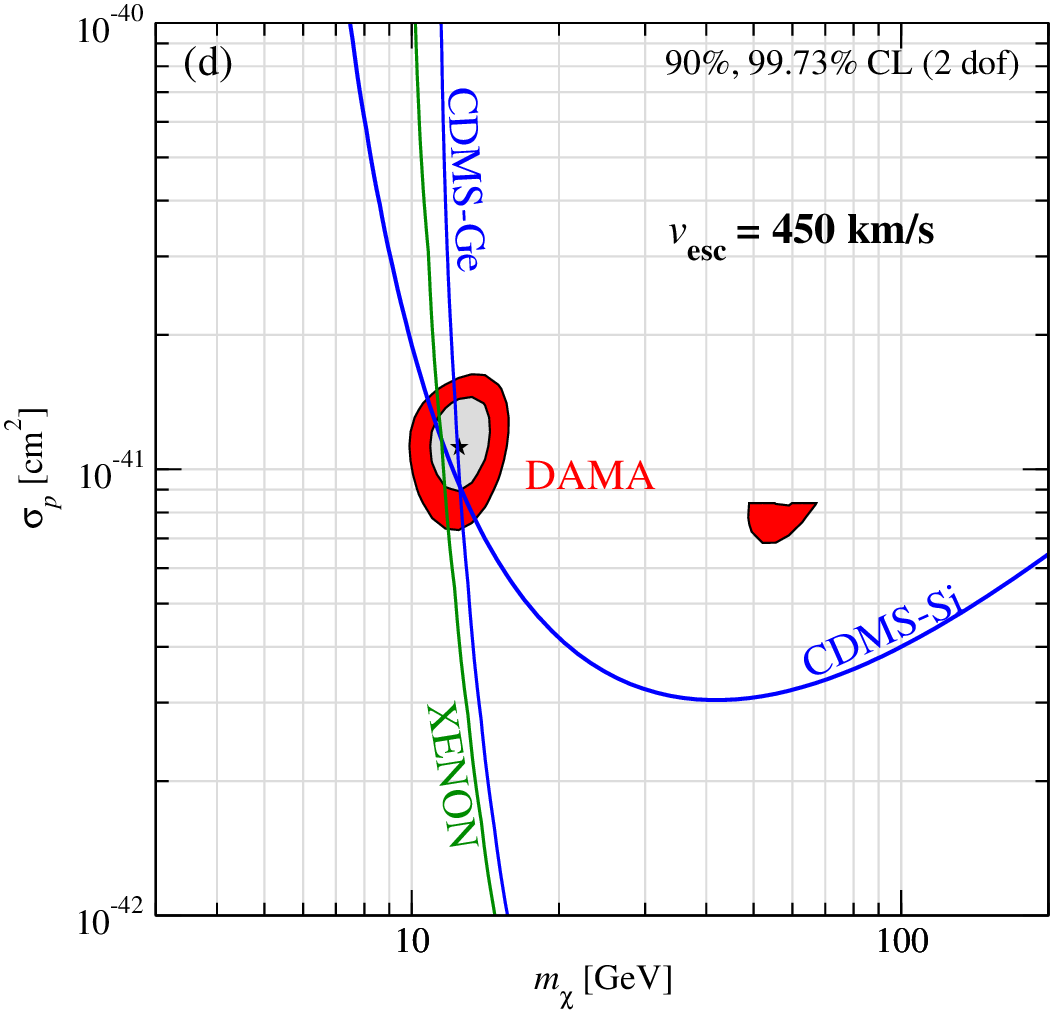} 
  \mycaption{\label{fig:regions-nonstd} Allowed regions at 90\% and
  99.73\%~CL for DAMA, and exclusion contours for CDMS-Si, CDMS-Ge and
  XENON at 90\%~CL for the DM halo obtained in the Via Lactea
  simulation (a), an isotropic Maxwellian halo with dispersion $\bar v
  = 110$~km/s (b), an asymmetric Maxwellian halo with dispersion $\bar
  v_R = 142$~km/s in the radial direction and $\bar v_T = 63$~km/s in
  the tangential direction (c), and an isotropic Maxwellian halo with
  dispersion $\bar v = 220$~km/s and escape velocity $v_\mathrm{esc} =
  450$~km/s (d). The best fit for DAMA is marked with a star. In the
  panels (b) and (c) we show also the 90\% and 99.73\%~CL regions
  for the global data combining all experiments, as well as the global
  best fit point (marked with a dot). }
\end{figure}

Using the data from the Via Lactea simulation, we have fitted for
values of $f_i$ and $\alpha_i$ as a function of radius from the centre
of the galaxy.  The results, which can be seen in figure \ref{fig:vl},
show that there is a considerable deviation from Gaussianity in the
velocity dispersion of the galaxy.  Both the deviation from
Gaussianity, the anisotropy of the velocity dispersion and the change
in the relationship between the width of the dispersion and the local
Keplerian velocity will change the ratio between the expected
modulation in the DAMA experiment and the total expected events at
XENON and CDMS.  We have calculated these changes and
attempted to see if they can increase the likelihood of the results
from both experiments being compatible, the results of the new fits
can be seen in figure~\ref{fig:regions-nonstd}(a).

It turns out that the deviation from the assumptions of the isothermal
sphere which are predicted by the Via Lactea simulation are not
sufficient to bring the region in parameter space favoured by DAMA
away from the region disfavoured by XENON and CDMS.  The numbers
associated with these regions are provided in table~\ref{tab:fits} and
show that using the velocity dispersion predicted by Via Lactea leads
to only a very small reduction in $\chi^2$ and the goodness of fit is
still unacceptably small.

\begin{table}
\begin{tabular}{l|cc|cc@{\quad}c@{\quad}cc}
\hline\hline
halo model & $\chi^2_\mathrm{DAMA,min}$ & $m_{\chi,\rm best}^{\rm DAMA}$ &
             $\chi^2_\mathrm{glob,min}$ & GOF &
             $\chi^2_\mathrm{PG}$ & PG & $m_{\chi,\rm best}^{\rm glob}$ \\
\hline
default analysis    & 36.8 & 12 & 59.3 & 0.05 & 22.6 & $1\times 10^{-5}$ & 9.5 \\
Via Lactea simulation
                    & 35.1 & 16 & 56.7 & 0.08 & 21.6 & $2\times 10^{-5}$ & 13.9  \\
Maxwellian 
$\bar v = 110$~km/s & 32.9 & 108& 46.8 & 0.32 & 13.5 & $1\times 10^{-3}$ & 16  \\
$\bar v_R = 142$~km/s, 
$\bar v_T = 63$~km/s& 32.7 & 18 & 39.6 & 0.62 & 6.5  & 0.04 & 18  \\
$v_\mathrm{esc} = 450$~km/s
                    & 36.5 & 12 & 51.6 & 0.17 & 15.1  & $5\times 10^{-4}$ & 11 \\
\hline\hline
\end{tabular}
  \mycaption{\label{tab:fits} Summary of the fits to DAMA data and
  global data (DAMA, CDMS-Ge, CDMS-Si, XENON) for different WIMP
  halos.  We give the best fit $\chi^2$ values, the goodness of fit
  (assuming 43~dof), the PG testing the consistency of DAMA with all
  other data, as well as the best fit WIMP masses (in GeV).}
\end{table}

It is therefore interesting to ask what kind of halo parameters could
lead to a better fit to the data, and how realistic would such
parameters be?  There are examples in the literature of the use of a
stream of dark matter to boost the annual modulation signal
\cite{Gondolo:2005hh}.  In this work, we choose to retain the
spherical symmetry of the halo density and instead of adding a stream, vary both the
width of the velocity dispersion and the anisotropy parameter $\beta_{vel}$
defined as
\begin{equation}
\beta_{vel}=1-\frac{\bar{v}_T^2}{\bar{v}_R^2} \,.
\end{equation}

A reduction in the width of the velocity
dispersion of dark matter alone helps reconcile the two data sets without the need to introduce anisotropy.  If
we assume an isotropic distribution of dark matter ($\beta_{vel}=0$)
and reduce $\bar{v}$ to 110 km/s which is half of the Keplerian
velocity at the solar radius, the goodness of fit increases
dramatically (see table~\ref{tab:fits}).  A further improvement in the
fit is made if one assumes a velocity dispersion lower than Keplerian,
but also highly anisotropic such that $\bar{v}_R=142$ km/s and
$\bar{v}_T=63$ km/s. As visible in figure~\ref{fig:regions-nonstd}~(c)
in this cases the entire 90\%~CL region of DAMA is consistent with the
90\%~CL bounds from CDMS and XENON. For the asymmetric velocity
distribution the global $\chi^2$ drops by about 20 units compared to
the default analysis and provides an excellent goodness of fit of
62\%.  The PG test gives compatibility of DAMA with NEV data with a
probability of 4\%, due to the remaining constraint from XENON.

A valid question is then whether such low and anisotropic values of
the velocity dispersions are at all realistic. In order to check on
the feasibility of such values, one needs to think about particular
dark matter halos and see if the solutions of the (Maxwell-) Jeans equations
allow simultaneously both a high velocity anisotropy ($\beta_{vel}
\sim 0.8$ ) and low velocity distribution at the location of the Sun.

In order to solve the Jeans equations, we will need to understand the distribution of mass in the galaxy.  Integration of a spherically symmetric dark matter profile with a well defined functional form is trivial.  However, at the radius of the Sun, it is important to consider not only dark
matter but also the presence of baryons, which make up most of the
mass in the central regions of the galaxy.  To model the Milky Way
baryon density we assume cylindrical symmetry and ignore any spiral
arms or bars. For the central bulge of stars we assume a density of
the form $\rho\propto r^{-\gamma}e^{-r/\lambda}$ while for the disk we
assume a (Kuzmin) delta function of matter in the $z$ direction ($z$
is the coordinate perpendicular to the disk) with a surface density
$\sigma_{\mathrm{disk}}(r) =
\frac{cM_{\mathrm{disk}\infty}}{2\pi\left(r^2+c^2\right)^{\frac{3}{2}}}$. We
choose the parameters of the model to match observations of the Milky
Way: $\gamma=1.85$, $\lambda=1 \,\mathrm{kpc}$, $c=5 \,\mathrm{kpc}$
and with the total disk and bulge mass $M_{\mathrm{disk}\infty} = 5
M_{\mathrm{bulge}}=6.5\times10^{10}M_{\odot}$ \cite{Zhao:1995qh,
Dehnen:1996fa, Klypin:2001xu, Kent:1991me}.  We assume that the disk
comes to an end at a radius of 15~kpc.

In order to parametrise our dark matter density profile we will
consider a profile which assumes two asymptotic radial power law
behaviors at both small ($\gamma$) and large ($\beta$) radii\footnote{This $\beta$ in the density profile should not be confused with the velocity dispersion anisotropy parameter $\beta_{vel}$.}. In this
profile, known as the '$\alpha\beta\gamma$' profile (or the Zhao
profile), the density as a function of radius is given by the
expression
\begin{equation}
\rho(r)=\frac{\rho_0}{(r/a)^\gamma\left[1+(r/a)^\alpha\right]^{\frac{\beta-\gamma}{\alpha}}}
\label{eq:abg}
\end{equation}
where $\alpha$ governs the radial rate at which the profile
interpolates between the asymptotic powers $-\gamma$ and $-\beta$.
The parameter $a$ is a characteristic scale radius determining the
location dividing the two regions described by a single power law.

Having assumed a value for $\alpha,\beta,\gamma$ and $a$ we then solve
for $\rho_0$ in order to get the correct value of the Keplerian
velocity at the solar radius.  This also determines the location of
the virial radius $r_{vir}$ which in this work is defined to be the
radius of the sphere within which the average density is 250 times the
critical density of the universe (we assume $h=0.7$).  The ratio
between $r_{vir}$ and $a$ is referred to as the concentration of the dark matter
halo.

Once we are in possession of these parameters, we can proceed to solve
the Jeans equation for the radial velocity dispersion
\cite{1987gady.book.....B}
\begin{equation}
\frac{1}{\rho}\frac{d\left(\rho\bar{v}_R^2\right)}{dr}+\frac{2\beta_{vel}\bar{v}_R^2}{r}=-\frac{d\phi}{dr}=-\frac{V_c^2}{r} \,,
\end{equation}
where $\phi(r)$ and $V_c(r)$ are the potential and Keplerian velocity
at a given radius.  We integrate this equation inwards from a large
radius several times the magnitude of $r_{vir}$ where we assume that
$\rho\bar{v}_R^2=0$.  We have checked that the result at $r=8.5$ kpc
is independent of the exact radius at which this boundary condition is
applied.  We have also assumed that, in the absence of a better
approximation, the anisotropy parameter $\beta_{vel}$ is a constant
throughout the halo.

\begin{figure}
\centering \includegraphics[height=0.5\textwidth]{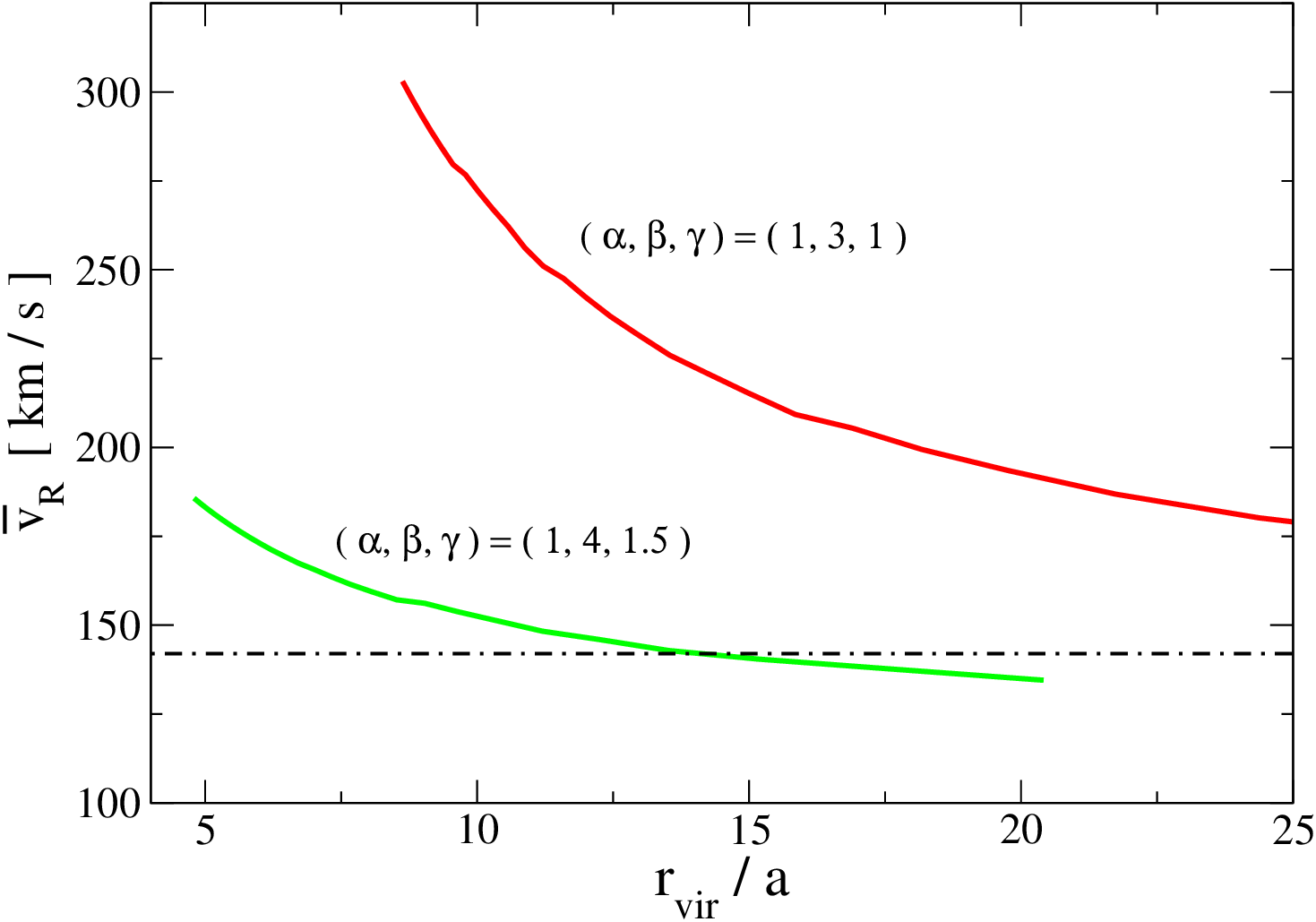}
  \mycaption{\label{fig:jeans} Here we plot the radial velocity
  dispersion $\bar v_R$ at the solar radius $r=8.5$~kpc as a function
  of the concentration of the dark matter halo $r_{vir}/a$ for two
  different dark matter profiles.  We have assumed that the velocity
  dispersion anisotropy parameter $\beta_{vel}=0.8$ and is a constant
  with respect to radius.  The horizontal line corresponds to the
  value of $\bar{v}_R$ which helps explain the discrepancy.  It
  appears that only for sets of halo parameters such as
  $(\alpha,\beta,\gamma)=(1,4,1.5)$ can one reconcile such a high
  value of $\beta$ with a low enough radial velocity dispersion to
  help explain the discrepancy between DAMA and XENON/CDMS.}
\end{figure}

The results are plotted in figure \ref{fig:jeans} and show that for a
NFW profile where the parameters are chosen such that
$(\alpha,\beta,\gamma)=(1,3,1)$ it seems to be rather difficult to
imagine that such a large value of the velocity anisotropy $\beta_{vel}$
could be consistent with low enough values of the velocity dispersion
to match the data.
We also look at a non-standard halo with
$(\alpha,\beta,\gamma)=(1,4,1.5)$.  The inner slope of such a halo is
quite steep, but even larger values of $\gamma$ than this may be
expected in dark matter halos where adiabatic contraction due to the
presence of baryons has occurred \cite{Gustafsson:2006gr}.  The rate
at which density decreases at larger radii is also larger than what is normally assumed.

If we are willing to accept such parameters for the dark matter
profile, it seems that the highly anisotropic value of $\beta_{vel}\sim 0.8$
that we require as one ingredient to make the DAMA data more
consistent with XENON and CDMS is not completely inconsistent with the
very low velocity dispersions that form the other ingredient.  The
analysis presented here is meant only as a suggestion of the magnitude
of possible effects.  If such explanations of the DAMA data were to be
taken seriously, a much deeper analysis of the Jeans equations should
be undertaken.  

Finally we mention that if one assumes a very low dark
matter escape velocity at the solar radius then one would remove many
of the fastest moving dark matter particles which would leave the
halo.  This would also result in more accord between DAMA and other
experiments but obtaining a large enough effect is difficult -- as
expressed in table~\ref{tab:fits}, lowering the escape velocity to 450~km/s
would only marginally make the fit more acceptable.  This, however,
must be considered an unrealistic solution, since the escape velocity
at the Solar radius is already 440~km/s even if there were no more
matter in the Galaxy at larger radii.

To summarise this section, it seems that without the use of streams
but rather by considering highly anisotropic velocity dispersions with
magnitudes far below the local Keplerian velocity at the radius of the
Sun would it be possible to reduce the conflict between DAMA and
XENON/CDMS.

\section{Conclusions}
\label{sec:conclusions}

Prompted by recent results from DAMA/LIBRA which establish the annual
modulation of their event rate at the 8.2$\sigma$ level, we have
studied the interpretation of this signal in terms of spin-independent
elastic WIMP scattering. We have shown that the energy spectrum of the
modulation signal strongly restricts the region of WIMP masses below
10~GeV, confining WIMP masses consistent with the DAMA data close to
$m_\chi \simeq 12$~GeV. This region is excluded by the limits from
CDMS and XENON, and therefore we conclude that even if channeling is
taken into account this interpretation of the DAMA modulation signal
is disfavoured. Applying a stringent test to evaluate the
consistency of DAMA with null-result experiments we find consistency
only with a formal probability of $10^{-5}$.

We have studied how robust this result is with respect to variations
of the WIMP velocity distribution in our galaxy by changing various
parameters of the distribution function. We find that decreasing the
dispersion of the distribution can somewhat reduce the tension in
the fit. Adopting in addition an asymmetric WIMP velocity profile with
a larger dispersion in the radial direction than tangential improves
the fit considerably.  We conclude that in principle it is possible to
reconsile DAMA in the considered framework, at the price of rather
exotic properties of the DM halo. The question remains whether such
halo properties can be realistic at all. We have checked that a WIMP
velocity distribution based on the Via Lactea $N$-body dark matter
simulation does not improve the fit considerably with respect to the
standard Maxwellian halo model.

Finally we mention that the negative conclusion on the compatibility
of DAMA with CDMS and XENON relies crucially on the energy threshold
of the latter two. In particular, a shift in the nuclear recoil energy
scale in these experiments may change the conclusion. Indeed, the new
measurements of the $\mathcal{L}_\mathrm{eff}$ parameter in
XENON~\cite{Sorensen:2008ec} (which has not been implemented in the
first arXiv version of this work) made the disagreement between DAMA
and XENON somewhat less sever.

\acknowledgments

We thank Graciela Gelmini for discussions in the initial stage of this
work and are very grateful to J\"urg Diemand for providing us with the
Via Lactea data. We thank the anonymous referee for pointing out the
new $\mathcal{L}_\mathrm{eff}$ measurement relevant for the XENON10
data analysis to us, and we acknowledge Laura Baudis for useful
correspondence on this issue.

\appendix

\section{Comments on the DAMA spectral information}
\label{sec:dama-mod}

\begin{figure}
\centering 
\includegraphics[height=0.4\textwidth]{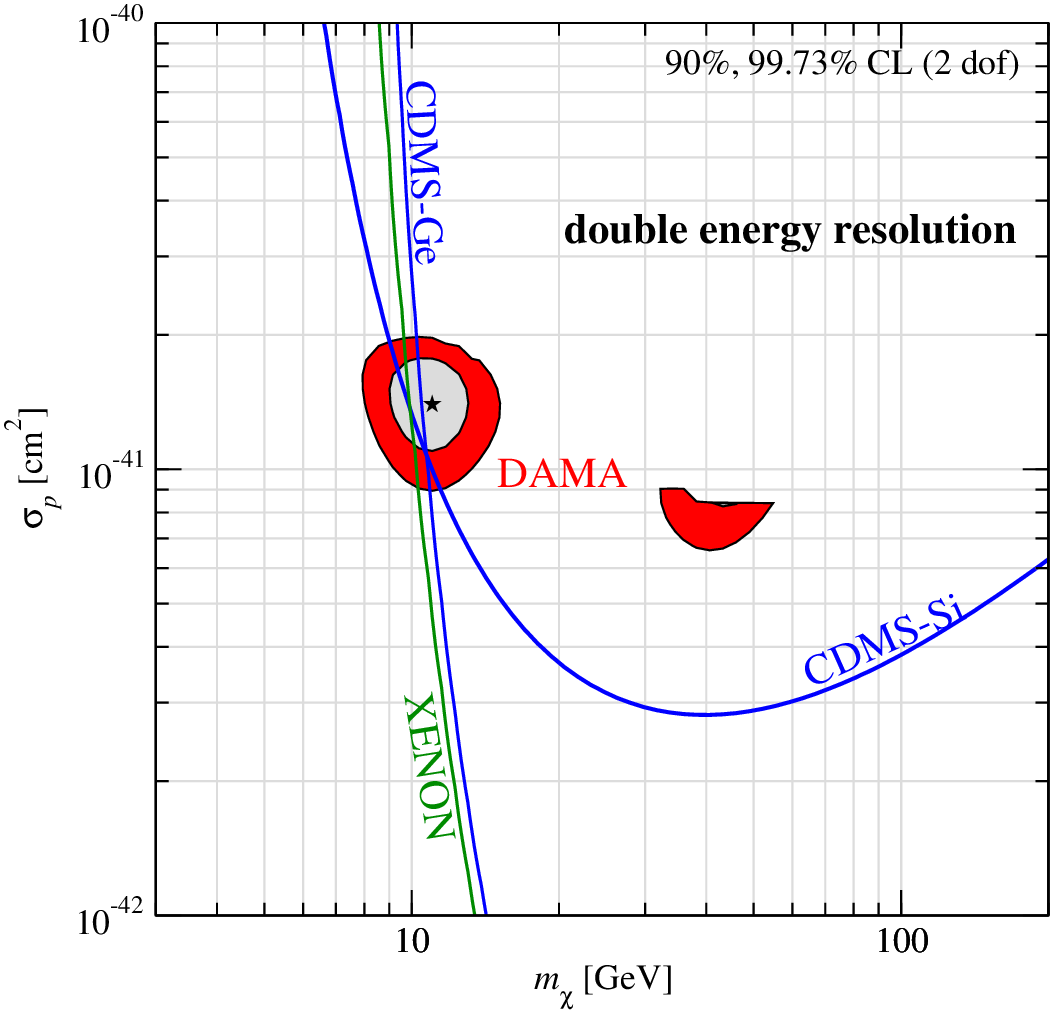} \qquad
\includegraphics[height=0.4\textwidth]{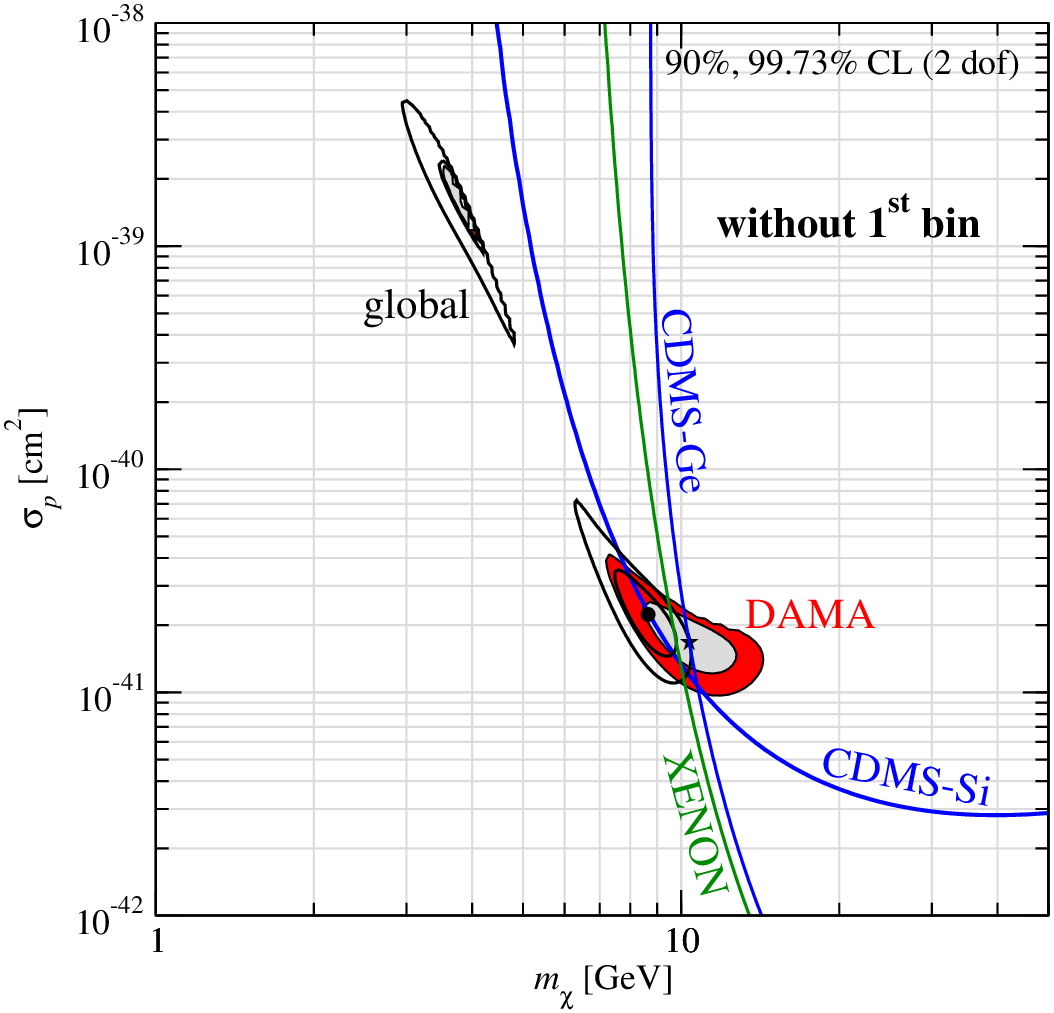}
  \mycaption{\label{fig:regions-mod-DAMA} Allowed regions at 90\% and
  99.73\%~CL for DAMA, and exclusion contours for CDMS-Si, CDMS-Ge and
  XENON at 90\%~CL for two ad-hoc modifications of the DAMA
  analysis. Left: we artificially assume an energy resolution in DAMA
  a factor two worse than the value given
  in~\cite{Bernabei:2008yh}. Right: omitting the lowest energy bin of
  the annual modulation spectrum between 2 and 2.5~keVee.  The best
  fit for DAMA is marked with a star. In the right panel we show also
  the 90\% and 99.73\%~CL regions for the global data combining all
  experiments, as well as the global best fit point (marked with a
  dot). }
\end{figure}

Our results are largely based on the fact that DAMA spectral
information excludes the low-mass WIMP region below
10~GeV. Obviously any effect which affects the spectral shape of the
signal will have an impact on this conclusion. First, the smearing due
to the energy resolution of the detector is important. We have checked
this by artificially increasing the width of the energy resolution
function given in Eq.~\ref{eq:Eres}~\cite{Bernabei:2008yh} by a factor
of two. The global fit improves by roughly 7 units in $\chi^2$, but
the tension between DAMA and NEV data persists at the level of
$8\times 10^{-4}$, compare table~\ref{tab:fits-mod} and
figure~\ref{fig:regions-mod-DAMA} (left).

\begin{table}
\begin{tabular}{l|cc|cc@{\quad}c@{\quad}cc}
\hline\hline
           & $\chi^2_\mathrm{DAMA,min}$ & $m_{\chi,\rm best}^{\rm DAMA}$ &
             $\chi^2_\mathrm{glob,min}$ & GOF &
             $\chi^2_\mathrm{PG}$ & PG & $m_{\chi,\rm best}^{\rm glob}$ \\
\hline
default analysis    & 36.8 & 12 & 59.3 & 0.05 & 22.6 & $1\times 10^{-5}$ & 9.5 \\
\hline
double $\sigma_E^\mathrm{DAMA}$ 
                    & 37.9 & 11 & 52.1 & 0.16 & 14.3 & $8\times 10^{-4}$ & 9.0 \\
w/o $1^\mathrm{st}$ DAMA data point
                    & 30.3 & 10 & 44.4 & 0.37 & 14.1 &  $9\times 10^{-4}$ & 8.6 \\
\hline\hline
\end{tabular}
  \mycaption{\label{tab:fits-mod} Summary of fits to DAMA data and
  global data (DAMA, CDMS-Ge, CDMS-Si, XENON) for the two ad-hoc
  modifications of the DAMA analysis of
  figure~\ref{fig:regions-mod-DAMA}.  We give the best fit $\chi^2$
  values, the goodness of fit (assuming 42~dof for the last row and
  43~dof otherwise), the PG testing the consistency of DAMA with all
  other data, as well as the best fit WIMP masses (in GeV).}
\end{table}

From figure~\ref{fig:spectrum} (left) it follows that the somewhat low
data point in the first energy bin is very important in constraining
the WIMP mass. We have repeated the analysis by excluding this bin
from the fit, using only the data on the modulation signal above
2.5~keVee. In this case the DAMA allowed region extends to lower
values of the WIMP mass, and once the NEV data are added the globally
allowed region includes values of $m_\chi \sim 4$~GeV and $\sigma_p
\sim 10^{-39}$~cm$^2$ at 90\%~CL. This region originates from
channeled events on Sodium which now can accommodate the spectrum
without the first bin, despite the contribution of un-channeled Na
events. Let us note, however, that in this region constraints from
other experiments, like CRESST-I~\cite{Altmann:2001ax},
TEXONO~\cite{Lin:2007ka}, or CoGeNT~\cite{Aalseth:2008rx} are
relevant. The global best fit point has an excellent
$\chi^2_\mathrm{glob,min} = 44.4/42$~dof.\footnote{The improvement of
  the consistency of NEV and DAMA data is only partially visible in
  the PG value of about 0.1\%, which is still rather low.  The reason
  is that also the fit of DAMA alone improves from
  $\chi^2_\mathrm{DAMA,min} = 36.8$ to 30.3 by dropping the first bin,
  which compensates partially the improvement in the global fit.}

Furthermore, we remark that any systematical uncertainty affecting the
low energy spectrum may be relevant. For example, figure~26
of~\cite{Bernabei:2008yh} shows that the efficiency for DAMA
single-hit events starts to deviate from 1 below about 8~keVee, just
in the signal region. Therefore, a possible uncertainty on this low
energy efficiency may affect the exclusion of the light WIMP
region. In the absence of detailed information on possible
energy-dependent systematic uncertainties we neglect such effects in
our analysis.
Let us note that the {\it ratio} of the signals in June and December
would be less affected by systematics, since any multiplicative
uncertainty (even energy dependent) would cancel, whereas the rate
{\it difference} published by the DAMA collaboration is affected by
such uncertainties.

Finally, we mention that the so-called Migdal effect could lead to
modifications of the predicted energy spectrum in DAMA,
see~\cite{Bernabei:2007jz} for a discussion and references. An
investigation of this effect is beyond the scope of this work.

\section{Comparison with other studies}
\label{app:comparison}

In this appendix we compare the results of our work to some studies
from other authors. The authors of~\cite{Petriello:2008jj} come to
a positive conclusion on the consistency between DAMA and constraints
from other experiments, since they do not include the information on
the spectral shape of the DAMA signal. In a work which appeared after
ours on the preprint server \cite{Hooper:2008cf}, the same authors
performed also an analysis including the spectrum which is in
agreement with our results.

Our work appeared on the preprint server basically at the same time as
\cite{Chang:2008xa}, with similar results. As in our study
these authors emphasize the importance of the spectral information of
the DAMA annual modulation and the constraint from the total
unmodulated rate. Whereas~\cite{Chang:2008xa} discusses DM streams,
our work considers non-standard halo models.

A similar study has been performed in \cite{Savage:2008er}, stressing
also the relevance of spectral information and extending the analysis
to spin-dependent cross sections. The general results for the
spin-independent case are in quantitative agreement with us,
though in some cases the authors draw different conclusions. In
particular, they use a variety of statistical tests complementary to
ours. Whereas our methods are largely based on parameter estimation
($\Delta\chi^2$ values with respect to the best fit point), these
authors show also contours of probabilities from a goodness-of-fit test
based on absolute $\chi^2$ values. As mentioned at the end of
section~\ref{sec:std-halo} this method is often not very sensitive to
a tension between different data sets, see also
footnote~\ref{ft:gof}. Furthermore, by showing contours up to the
5 and even 7$\sigma$~CL they do find overlap regions. 
Let us also comment on the best fit point for DAMA, obtained at
$m_\chi = 80$~GeV in Tab.~IV of~\cite{Savage:2008er}, compared to our
result $m_\chi = 12$~GeV from Eq.~\ref{eq:dama-bfp}. The reason why
we disfavour the fit in the large DM mass region is the inclusion of
the constraint from the unmodulated rate in DAMA, which cuts away
large part of this region, including also the best fit point of
\cite{Savage:2008er}, compare figure~\ref{fig:regions}, and shifts the
global best fit point to the low mass region.

\bibliographystyle{apsrev}
\bibliography{./dm}

\begin{thebibliography}{53}
\expandafter\ifx\csname natexlab\endcsname\relax\def\natexlab#1{#1}\fi
\expandafter\ifx\csname bibnamefont\endcsname\relax
  \def\bibnamefont#1{#1}\fi
\expandafter\ifx\csname bibfnamefont\endcsname\relax
  \def\bibfnamefont#1{#1}\fi
\expandafter\ifx\csname citenamefont\endcsname\relax
  \def\citenamefont#1{#1}\fi
\expandafter\ifx\csname url\endcsname\relax
  \def\url#1{\texttt{#1}}\fi
\expandafter\ifx\csname urlprefix\endcsname\relax\def\urlprefix{URL }\fi
\providecommand{\bibinfo}[2]{#2}
\providecommand{\eprint}[2][]{\url{#2}}

\bibitem[{\citenamefont{Bernabei et~al.}(2008{\natexlab{a}})}]{Bernabei:2008yi}
\bibinfo{author}{\bibfnamefont{R.}~\bibnamefont{Bernabei}} \bibnamefont{et~al.}
  (\bibinfo{collaboration}{DAMA}) (\bibinfo{year}{2008}{\natexlab{a}}),
  \eprint{0804.2741}.

\bibitem[{\citenamefont{Jungman et~al.}(1996)\citenamefont{Jungman,
  Kamionkowski, and Griest}}]{Jungman:1995df}
\bibinfo{author}{\bibfnamefont{G.}~\bibnamefont{Jungman}},
  \bibinfo{author}{\bibfnamefont{M.}~\bibnamefont{Kamionkowski}},
  \bibnamefont{and} \bibinfo{author}{\bibfnamefont{K.}~\bibnamefont{Griest}},
  \bibinfo{journal}{Phys. Rept.} \textbf{\bibinfo{volume}{267}},
  \bibinfo{pages}{195} (\bibinfo{year}{1996}), \eprint{hep-ph/9506380}.

\bibitem[{\citenamefont{Ahmed et~al.}(2008)}]{Ahmed:2008eu}
\bibinfo{author}{\bibfnamefont{Z.}~\bibnamefont{Ahmed}} \bibnamefont{et~al.}
  (\bibinfo{collaboration}{CDMS}) (\bibinfo{year}{2008}), \eprint{0802.3530}.

\bibitem[{\citenamefont{Angle et~al.}(2008)}]{Angle:2007uj}
\bibinfo{author}{\bibfnamefont{J.}~\bibnamefont{Angle}} \bibnamefont{et~al.}
  (\bibinfo{collaboration}{XENON}), \bibinfo{journal}{Phys. Rev. Lett.}
  \textbf{\bibinfo{volume}{100}}, \bibinfo{pages}{021303}
  (\bibinfo{year}{2008}), \eprint{0706.0039}.

\bibitem[{\citenamefont{Ullio et~al.}(2001)\citenamefont{Ullio, Kamionkowski,
  and Vogel}}]{Ullio:2000bv}
\bibinfo{author}{\bibfnamefont{P.}~\bibnamefont{Ullio}},
  \bibinfo{author}{\bibfnamefont{M.}~\bibnamefont{Kamionkowski}},
  \bibnamefont{and} \bibinfo{author}{\bibfnamefont{P.}~\bibnamefont{Vogel}},
  \bibinfo{journal}{JHEP} \textbf{\bibinfo{volume}{07}}, \bibinfo{pages}{044}
  (\bibinfo{year}{2001}), \eprint{hep-ph/0010036}.

\bibitem[{\citenamefont{Savage et~al.}(2004)\citenamefont{Savage, Gondolo, and
  Freese}}]{Savage:2004fn}
\bibinfo{author}{\bibfnamefont{C.}~\bibnamefont{Savage}},
  \bibinfo{author}{\bibfnamefont{P.}~\bibnamefont{Gondolo}}, \bibnamefont{and}
  \bibinfo{author}{\bibfnamefont{K.}~\bibnamefont{Freese}},
  \bibinfo{journal}{Phys. Rev.} \textbf{\bibinfo{volume}{D70}},
  \bibinfo{pages}{123513} (\bibinfo{year}{2004}), \eprint{astro-ph/0408346}.

\bibitem[{\citenamefont{Bottino
  et~al.}(2003{\natexlab{a}})\citenamefont{Bottino, Donato, Fornengo, and
  Scopel}}]{Bottino:2003iu}
\bibinfo{author}{\bibfnamefont{A.}~\bibnamefont{Bottino}},
  \bibinfo{author}{\bibfnamefont{F.}~\bibnamefont{Donato}},
  \bibinfo{author}{\bibfnamefont{N.}~\bibnamefont{Fornengo}}, \bibnamefont{and}
  \bibinfo{author}{\bibfnamefont{S.}~\bibnamefont{Scopel}},
  \bibinfo{journal}{Phys. Rev.} \textbf{\bibinfo{volume}{D68}},
  \bibinfo{pages}{043506} (\bibinfo{year}{2003}{\natexlab{a}}),
  \eprint{hep-ph/0304080}.

\bibitem[{\citenamefont{Bottino et~al.}(2004)\citenamefont{Bottino, Donato,
  Fornengo, and Scopel}}]{Bottino:2003cz}
\bibinfo{author}{\bibfnamefont{A.}~\bibnamefont{Bottino}},
  \bibinfo{author}{\bibfnamefont{F.}~\bibnamefont{Donato}},
  \bibinfo{author}{\bibfnamefont{N.}~\bibnamefont{Fornengo}}, \bibnamefont{and}
  \bibinfo{author}{\bibfnamefont{S.}~\bibnamefont{Scopel}},
  \bibinfo{journal}{Phys. Rev.} \textbf{\bibinfo{volume}{D69}},
  \bibinfo{pages}{037302} (\bibinfo{year}{2004}), \eprint{hep-ph/0307303}.

\bibitem[{\citenamefont{Gondolo and Gelmini}(2005)}]{Gondolo:2005hh}
\bibinfo{author}{\bibfnamefont{P.}~\bibnamefont{Gondolo}} \bibnamefont{and}
  \bibinfo{author}{\bibfnamefont{G.}~\bibnamefont{Gelmini}},
  \bibinfo{journal}{Phys. Rev.} \textbf{\bibinfo{volume}{D71}},
  \bibinfo{pages}{123520} (\bibinfo{year}{2005}), \eprint{hep-ph/0504010}.

\bibitem[{\citenamefont{Bernabei et~al.}(2006)}]{Bernabei:2005ca}
\bibinfo{author}{\bibfnamefont{R.}~\bibnamefont{Bernabei}}
  \bibnamefont{et~al.}, \bibinfo{journal}{Int. J. Mod. Phys.}
  \textbf{\bibinfo{volume}{A21}}, \bibinfo{pages}{1445} (\bibinfo{year}{2006}),
  \eprint{astro-ph/0511262}.

\bibitem[{\citenamefont{Pospelov et~al.}(2008)\citenamefont{Pospelov, Ritz, and
  Voloshin}}]{Pospelov:2008jk}
\bibinfo{author}{\bibfnamefont{M.}~\bibnamefont{Pospelov}},
  \bibinfo{author}{\bibfnamefont{A.}~\bibnamefont{Ritz}}, \bibnamefont{and}
  \bibinfo{author}{\bibfnamefont{M.~B.} \bibnamefont{Voloshin}}
  (\bibinfo{year}{2008}), \eprint{0807.3279}.

\bibitem[{\citenamefont{Gondolo and Raffelt}(2008)}]{Gondolo:2008dd}
\bibinfo{author}{\bibfnamefont{P.}~\bibnamefont{Gondolo}} \bibnamefont{and}
  \bibinfo{author}{\bibfnamefont{G.}~\bibnamefont{Raffelt}}
  (\bibinfo{year}{2008}), \eprint{0807.2926}.

\bibitem[{\citenamefont{Bernabei et~al.}(2008{\natexlab{b}})}]{Bernabei:2007gr}
\bibinfo{author}{\bibfnamefont{R.}~\bibnamefont{Bernabei}}
  \bibnamefont{et~al.}, \bibinfo{journal}{Phys. Rev.}
  \textbf{\bibinfo{volume}{D77}}, \bibinfo{pages}{023506}
  (\bibinfo{year}{2008}{\natexlab{b}}), \eprint{0712.0562}.

\bibitem[{\citenamefont{Tucker-Smith and Weiner}(2001)}]{TuckerSmith:2001hy}
\bibinfo{author}{\bibfnamefont{D.}~\bibnamefont{Tucker-Smith}}
  \bibnamefont{and} \bibinfo{author}{\bibfnamefont{N.}~\bibnamefont{Weiner}},
  \bibinfo{journal}{Phys. Rev.} \textbf{\bibinfo{volume}{D64}},
  \bibinfo{pages}{043502} (\bibinfo{year}{2001}), \eprint{hep-ph/0101138}.

\bibitem[{\citenamefont{Chang et~al.}(2008{\natexlab{a}})\citenamefont{Chang,
  Kribs, Tucker-Smith, and Weiner}}]{Chang:2008gd}
\bibinfo{author}{\bibfnamefont{S.}~\bibnamefont{Chang}},
  \bibinfo{author}{\bibfnamefont{G.~D.} \bibnamefont{Kribs}},
  \bibinfo{author}{\bibfnamefont{D.}~\bibnamefont{Tucker-Smith}},
  \bibnamefont{and} \bibinfo{author}{\bibfnamefont{N.}~\bibnamefont{Weiner}}
  (\bibinfo{year}{2008}{\natexlab{a}}), \eprint{0807.2250}.

\bibitem[{\citenamefont{Foot}(2008)}]{Foot:2008nw}
\bibinfo{author}{\bibfnamefont{R.}~\bibnamefont{Foot}} (\bibinfo{year}{2008}),
  \eprint{0804.4518}.

\bibitem[{\citenamefont{Bottino
  et~al.}(2008{\natexlab{a}})\citenamefont{Bottino, Donato, Fornengo, and
  Scopel}}]{Bottino:2007qg}
\bibinfo{author}{\bibfnamefont{A.}~\bibnamefont{Bottino}},
  \bibinfo{author}{\bibfnamefont{F.}~\bibnamefont{Donato}},
  \bibinfo{author}{\bibfnamefont{N.}~\bibnamefont{Fornengo}}, \bibnamefont{and}
  \bibinfo{author}{\bibfnamefont{S.}~\bibnamefont{Scopel}},
  \bibinfo{journal}{Phys. Rev.} \textbf{\bibinfo{volume}{D77}},
  \bibinfo{pages}{015002} (\bibinfo{year}{2008}{\natexlab{a}}),
  \eprint{0710.0553}.

\bibitem[{\citenamefont{Bottino
  et~al.}(2008{\natexlab{b}})\citenamefont{Bottino, Donato, Fornengo, and
  Scopel}}]{Bottino:2008mf}
\bibinfo{author}{\bibfnamefont{A.}~\bibnamefont{Bottino}},
  \bibinfo{author}{\bibfnamefont{F.}~\bibnamefont{Donato}},
  \bibinfo{author}{\bibfnamefont{N.}~\bibnamefont{Fornengo}}, \bibnamefont{and}
  \bibinfo{author}{\bibfnamefont{S.}~\bibnamefont{Scopel}}
  (\bibinfo{year}{2008}{\natexlab{b}}), \eprint{0806.4099}.

\bibitem[{\citenamefont{Petriello and Zurek}(2008)}]{Petriello:2008jj}
\bibinfo{author}{\bibfnamefont{F.}~\bibnamefont{Petriello}} \bibnamefont{and}
  \bibinfo{author}{\bibfnamefont{K.~M.} \bibnamefont{Zurek}}
  (\bibinfo{year}{2008}), \eprint{0806.3989}.

\bibitem[{\citenamefont{Feng et~al.}(2008)\citenamefont{Feng, Kumar, and
  Strigari}}]{Feng:2008dz}
\bibinfo{author}{\bibfnamefont{J.~L.} \bibnamefont{Feng}},
  \bibinfo{author}{\bibfnamefont{J.}~\bibnamefont{Kumar}}, \bibnamefont{and}
  \bibinfo{author}{\bibfnamefont{L.~E.} \bibnamefont{Strigari}}
  (\bibinfo{year}{2008}), \eprint{0806.3746}.

\bibitem[{\citenamefont{Drobyshevski}(2007)}]{Drobyshevski:2007zj}
\bibinfo{author}{\bibfnamefont{E.~M.} \bibnamefont{Drobyshevski}}
  (\bibinfo{year}{2007}), \eprint{0706.3095}.

\bibitem[{\citenamefont{Bottino
  et~al.}(2003{\natexlab{b}})\citenamefont{Bottino, Fornengo, and
  Scopel}}]{Bottino:2002ry}
\bibinfo{author}{\bibfnamefont{A.}~\bibnamefont{Bottino}},
  \bibinfo{author}{\bibfnamefont{N.}~\bibnamefont{Fornengo}}, \bibnamefont{and}
  \bibinfo{author}{\bibfnamefont{S.}~\bibnamefont{Scopel}},
  \bibinfo{journal}{Phys. Rev.} \textbf{\bibinfo{volume}{D67}},
  \bibinfo{pages}{063519} (\bibinfo{year}{2003}{\natexlab{b}}),
  \eprint{hep-ph/0212379}.

\bibitem[{\citenamefont{Barger et~al.}(2005)\citenamefont{Barger, Langacker,
  and Lee}}]{Barger:2005hb}
\bibinfo{author}{\bibfnamefont{V.}~\bibnamefont{Barger}},
  \bibinfo{author}{\bibfnamefont{P.}~\bibnamefont{Langacker}},
  \bibnamefont{and} \bibinfo{author}{\bibfnamefont{H.-S.} \bibnamefont{Lee}},
  \bibinfo{journal}{Phys. Lett.} \textbf{\bibinfo{volume}{B630}},
  \bibinfo{pages}{85} (\bibinfo{year}{2005}), \eprint{hep-ph/0508027}.

\bibitem[{\citenamefont{Gunion et~al.}(2006)\citenamefont{Gunion, Hooper, and
  McElrath}}]{Gunion:2005rw}
\bibinfo{author}{\bibfnamefont{J.~F.} \bibnamefont{Gunion}},
  \bibinfo{author}{\bibfnamefont{D.}~\bibnamefont{Hooper}}, \bibnamefont{and}
  \bibinfo{author}{\bibfnamefont{B.}~\bibnamefont{McElrath}},
  \bibinfo{journal}{Phys. Rev.} \textbf{\bibinfo{volume}{D73}},
  \bibinfo{pages}{015011} (\bibinfo{year}{2006}), \eprint{hep-ph/0509024}.

\bibitem[{\citenamefont{Altmann et~al.}(2001)}]{Altmann:2001ax}
\bibinfo{author}{\bibfnamefont{M.~F.} \bibnamefont{Altmann}}
  \bibnamefont{et~al.} (\bibinfo{collaboration}{CRESST-I})
  (\bibinfo{year}{2001}), \eprint{astro-ph/0106314}.

\bibitem[{\citenamefont{Akerib et~al.}(2003)}]{Akerib:2003px}
\bibinfo{author}{\bibfnamefont{D.~S.} \bibnamefont{Akerib}}
  \bibnamefont{et~al.} (\bibinfo{collaboration}{CDMS}), \bibinfo{journal}{Phys.
  Rev.} \textbf{\bibinfo{volume}{D68}}, \bibinfo{pages}{082002}
  (\bibinfo{year}{2003}), \eprint{hep-ex/0306001}.

\bibitem[{\citenamefont{Lin et~al.}(2007)}]{Lin:2007ka}
\bibinfo{author}{\bibfnamefont{S.~T.} \bibnamefont{Lin}} \bibnamefont{et~al.}
  (\bibinfo{collaboration}{TEXONO}) (\bibinfo{year}{2007}), \eprint{0712.1645}.

\bibitem[{\citenamefont{Aalseth et~al.}(2008)}]{Aalseth:2008rx}
\bibinfo{author}{\bibfnamefont{C.~E.} \bibnamefont{Aalseth}}
  \bibnamefont{et~al.} (\bibinfo{collaboration}{CoGeNT})
  (\bibinfo{year}{2008}), \eprint{0807.0879}.

\bibitem[{\citenamefont{Akerib et~al.}(2006)}]{Akerib:2005kh}
\bibinfo{author}{\bibfnamefont{D.~S.} \bibnamefont{Akerib}}
  \bibnamefont{et~al.} (\bibinfo{collaboration}{CDMS}), \bibinfo{journal}{Phys.
  Rev. Lett.} \textbf{\bibinfo{volume}{96}}, \bibinfo{pages}{011302}
  (\bibinfo{year}{2006}), \eprint{astro-ph/0509259}.

\bibitem[{\citenamefont{Belli et~al.}(2002)\citenamefont{Belli, Cerulli,
  Fornengo, and Scopel}}]{Belli:2002yt}
\bibinfo{author}{\bibfnamefont{P.}~\bibnamefont{Belli}},
  \bibinfo{author}{\bibfnamefont{R.}~\bibnamefont{Cerulli}},
  \bibinfo{author}{\bibfnamefont{N.}~\bibnamefont{Fornengo}}, \bibnamefont{and}
  \bibinfo{author}{\bibfnamefont{S.}~\bibnamefont{Scopel}},
  \bibinfo{journal}{Phys. Rev.} \textbf{\bibinfo{volume}{D66}},
  \bibinfo{pages}{043503} (\bibinfo{year}{2002}), \eprint{hep-ph/0203242}.

\bibitem[{\citenamefont{Fornengo and Scopel}(2003)}]{Fornengo:2003fm}
\bibinfo{author}{\bibfnamefont{N.}~\bibnamefont{Fornengo}} \bibnamefont{and}
  \bibinfo{author}{\bibfnamefont{S.}~\bibnamefont{Scopel}},
  \bibinfo{journal}{Phys. Lett.} \textbf{\bibinfo{volume}{B576}},
  \bibinfo{pages}{189} (\bibinfo{year}{2003}), \eprint{hep-ph/0301132}.

\bibitem[{\citenamefont{Green}(2002)}]{Green:2002ht}
\bibinfo{author}{\bibfnamefont{A.~M.} \bibnamefont{Green}},
  \bibinfo{journal}{Phys. Rev.} \textbf{\bibinfo{volume}{D66}},
  \bibinfo{pages}{083003} (\bibinfo{year}{2002}), \eprint{astro-ph/0207366}.

\bibitem[{\citenamefont{Vergados et~al.}(2008)\citenamefont{Vergados, Hansen,
  and Host}}]{Vergados:2007nc}
\bibinfo{author}{\bibfnamefont{J.~D.} \bibnamefont{Vergados}},
  \bibinfo{author}{\bibfnamefont{S.~H.} \bibnamefont{Hansen}},
  \bibnamefont{and} \bibinfo{author}{\bibfnamefont{O.}~\bibnamefont{Host}},
  \bibinfo{journal}{Phys. Rev.} \textbf{\bibinfo{volume}{D77}},
  \bibinfo{pages}{023509} (\bibinfo{year}{2008}), \eprint{0711.4895}.

\bibitem[{\citenamefont{Diemand et~al.}(2007)\citenamefont{Diemand, Kuhlen, and
  Madau}}]{Diemand:2006ik}
\bibinfo{author}{\bibfnamefont{J.}~\bibnamefont{Diemand}},
  \bibinfo{author}{\bibfnamefont{M.}~\bibnamefont{Kuhlen}}, \bibnamefont{and}
  \bibinfo{author}{\bibfnamefont{P.}~\bibnamefont{Madau}},
  \bibinfo{journal}{Astrophys. J.} \textbf{\bibinfo{volume}{657}},
  \bibinfo{pages}{262} (\bibinfo{year}{2007}), \eprint{astro-ph/0611370}.

\bibitem[{\citenamefont{Gelmini and Gondolo}(2001)}]{Gelmini:2000dm}
\bibinfo{author}{\bibfnamefont{G.}~\bibnamefont{Gelmini}} \bibnamefont{and}
  \bibinfo{author}{\bibfnamefont{P.}~\bibnamefont{Gondolo}},
  \bibinfo{journal}{Phys. Rev.} \textbf{\bibinfo{volume}{D64}},
  \bibinfo{pages}{023504} (\bibinfo{year}{2001}), \eprint{hep-ph/0012315}.

\bibitem[{\citenamefont{Yao et~al.}(2006)}]{pdg}
\bibinfo{author}{\bibfnamefont{W.-M.} \bibnamefont{Yao}} \bibnamefont{et~al.}
  (\bibinfo{collaboration}{Particle Data Group}), \bibinfo{journal}{J. Phys. G}
  \textbf{\bibinfo{volume}{33}}, \bibinfo{pages}{1} (\bibinfo{year}{2006}).

\bibitem[{\citenamefont{Green}(2003)}]{Green:2003yh}
\bibinfo{author}{\bibfnamefont{A.~M.} \bibnamefont{Green}},
  \bibinfo{journal}{Phys. Rev.} \textbf{\bibinfo{volume}{D68}},
  \bibinfo{pages}{023004} (\bibinfo{year}{2003}), \eprint{astro-ph/0304446}.

\bibitem[{\citenamefont{Bernabei et~al.}(2008{\natexlab{c}})}]{Bernabei:2007hw}
\bibinfo{author}{\bibfnamefont{R.}~\bibnamefont{Bernabei}}
  \bibnamefont{et~al.}, \bibinfo{journal}{Eur. Phys. J.}
  \textbf{\bibinfo{volume}{C53}}, \bibinfo{pages}{205}
  (\bibinfo{year}{2008}{\natexlab{c}}), \eprint{0710.0288}.

\bibitem[{\citenamefont{Bernabei et~al.}(2008{\natexlab{d}})}]{Bernabei:2008yh}
\bibinfo{author}{\bibfnamefont{R.}~\bibnamefont{Bernabei}} \bibnamefont{et~al.}
  (\bibinfo{collaboration}{DAMA}), \bibinfo{journal}{Nucl. Instrum. Meth.}
  \textbf{\bibinfo{volume}{A592}}, \bibinfo{pages}{297}
  (\bibinfo{year}{2008}{\natexlab{d}}), \eprint{0804.2738}.

\bibitem[{\citenamefont{Sorensen et~al.}(2008)}]{Sorensen:2008ec}
\bibinfo{author}{\bibfnamefont{P.}~\bibnamefont{Sorensen}} \bibnamefont{et~al.}
  (\bibinfo{year}{2008}), \eprint{0807.0459}.

\bibitem[{\citenamefont{Maltoni and Schwetz}(2003)}]{Maltoni:2003cu}
\bibinfo{author}{\bibfnamefont{M.}~\bibnamefont{Maltoni}} \bibnamefont{and}
  \bibinfo{author}{\bibfnamefont{T.}~\bibnamefont{Schwetz}},
  \bibinfo{journal}{Phys. Rev.} \textbf{\bibinfo{volume}{D68}},
  \bibinfo{pages}{033020} (\bibinfo{year}{2003}), \eprint{hep-ph/0304176}.

\bibitem[{\citenamefont{Springel et~al.}(2005)}]{Springel:2005nw}
\bibinfo{author}{\bibfnamefont{V.}~\bibnamefont{Springel}}
  \bibnamefont{et~al.}, \bibinfo{journal}{Nature}
  \textbf{\bibinfo{volume}{435}}, \bibinfo{pages}{629} (\bibinfo{year}{2005}),
  \eprint{astro-ph/0504097}.

\bibitem[{\citenamefont{Navarro et~al.}(1996)\citenamefont{Navarro, Frenk, and
  White}}]{Navarro:1995iw}
\bibinfo{author}{\bibfnamefont{J.~F.} \bibnamefont{Navarro}},
  \bibinfo{author}{\bibfnamefont{C.~S.} \bibnamefont{Frenk}}, \bibnamefont{and}
  \bibinfo{author}{\bibfnamefont{S.~D.~M.} \bibnamefont{White}},
  \bibinfo{journal}{Astrophys. J.} \textbf{\bibinfo{volume}{462}},
  \bibinfo{pages}{563} (\bibinfo{year}{1996}), \eprint{astro-ph/9508025}.

\bibitem[{\citenamefont{Gustafsson et~al.}(2006)\citenamefont{Gustafsson,
  Fairbairn, and Sommer-Larsen}}]{Gustafsson:2006gr}
\bibinfo{author}{\bibfnamefont{M.}~\bibnamefont{Gustafsson}},
  \bibinfo{author}{\bibfnamefont{M.}~\bibnamefont{Fairbairn}},
  \bibnamefont{and}
  \bibinfo{author}{\bibfnamefont{J.}~\bibnamefont{Sommer-Larsen}},
  \bibinfo{journal}{Phys. Rev.} \textbf{\bibinfo{volume}{D74}},
  \bibinfo{pages}{123522} (\bibinfo{year}{2006}), \eprint{astro-ph/0608634}.

\bibitem[{\citenamefont{Zhao}(1995)}]{Zhao:1995qh}
\bibinfo{author}{\bibfnamefont{H.}~\bibnamefont{Zhao}} (\bibinfo{year}{1995}),
  \eprint{astro-ph/9512064}.

\bibitem[{\citenamefont{Dehnen and Binney}(1998)}]{Dehnen:1996fa}
\bibinfo{author}{\bibfnamefont{W.}~\bibnamefont{Dehnen}} \bibnamefont{and}
  \bibinfo{author}{\bibfnamefont{J.}~\bibnamefont{Binney}},
  \bibinfo{journal}{Mon. Not. Roy. Astron. Soc.}
  \textbf{\bibinfo{volume}{294}}, \bibinfo{pages}{429} (\bibinfo{year}{1998}),
  \eprint{astro-ph/9612059}.

\bibitem[{\citenamefont{Klypin et~al.}(2002)\citenamefont{Klypin, Zhao, and
  Somerville}}]{Klypin:2001xu}
\bibinfo{author}{\bibfnamefont{A.}~\bibnamefont{Klypin}},
  \bibinfo{author}{\bibfnamefont{H.}~\bibnamefont{Zhao}}, \bibnamefont{and}
  \bibinfo{author}{\bibfnamefont{R.~S.} \bibnamefont{Somerville}},
  \bibinfo{journal}{Astrophys. J.} \textbf{\bibinfo{volume}{573}},
  \bibinfo{pages}{597} (\bibinfo{year}{2002}), \eprint{astro-ph/0110390}.

\bibitem[{\citenamefont{Kent et~al.}(1991)\citenamefont{Kent, Dame, and
  Fazio}}]{Kent:1991me}
\bibinfo{author}{\bibfnamefont{S.~M.} \bibnamefont{Kent}},
  \bibinfo{author}{\bibfnamefont{T.~M.} \bibnamefont{Dame}}, \bibnamefont{and}
  \bibinfo{author}{\bibfnamefont{G.}~\bibnamefont{Fazio}},
  \bibinfo{journal}{Astrophys. J.} \textbf{\bibinfo{volume}{378}},
  \bibinfo{pages}{131} (\bibinfo{year}{1991}).

\bibitem[{\citenamefont{{Binney} and {Tremaine}}(1987)}]{1987gady.book.....B}
\bibinfo{author}{\bibfnamefont{J.}~\bibnamefont{{Binney}}} \bibnamefont{and}
  \bibinfo{author}{\bibfnamefont{S.}~\bibnamefont{{Tremaine}}}
  (\bibinfo{year}{1987}), \bibinfo{note}{{"Galactic dynamics", Princeton
  University Press.}}

\bibitem[{\citenamefont{Bernabei et~al.}(2007)}]{Bernabei:2007jz}
\bibinfo{author}{\bibfnamefont{R.}~\bibnamefont{Bernabei}}
  \bibnamefont{et~al.}, \bibinfo{journal}{Int. J. Mod. Phys.}
  \textbf{\bibinfo{volume}{A22}}, \bibinfo{pages}{3155} (\bibinfo{year}{2007}),
  \eprint{0706.1421}.

\bibitem[{\citenamefont{Hooper et~al.}(2008)\citenamefont{Hooper, Petriello,
  Zurek, and Kamionkowski}}]{Hooper:2008cf}
\bibinfo{author}{\bibfnamefont{D.}~\bibnamefont{Hooper}},
  \bibinfo{author}{\bibfnamefont{F.}~\bibnamefont{Petriello}},
  \bibinfo{author}{\bibfnamefont{K.~M.} \bibnamefont{Zurek}}, \bibnamefont{and}
  \bibinfo{author}{\bibfnamefont{M.}~\bibnamefont{Kamionkowski}}
  (\bibinfo{year}{2008}), \eprint{0808.2464}.

\bibitem[{\citenamefont{Chang et~al.}(2008{\natexlab{b}})\citenamefont{Chang,
  Pierce, and Weiner}}]{Chang:2008xa}
\bibinfo{author}{\bibfnamefont{S.}~\bibnamefont{Chang}},
  \bibinfo{author}{\bibfnamefont{A.}~\bibnamefont{Pierce}}, \bibnamefont{and}
  \bibinfo{author}{\bibfnamefont{N.}~\bibnamefont{Weiner}}
  (\bibinfo{year}{2008}{\natexlab{b}}), \eprint{0808.0196}.

\bibitem[{\citenamefont{Savage et~al.}(2008)\citenamefont{Savage, Gelmini,
  Gondolo, and Freese}}]{Savage:2008er}
\bibinfo{author}{\bibfnamefont{C.}~\bibnamefont{Savage}},
  \bibinfo{author}{\bibfnamefont{G.}~\bibnamefont{Gelmini}},
  \bibinfo{author}{\bibfnamefont{P.}~\bibnamefont{Gondolo}}, \bibnamefont{and}
  \bibinfo{author}{\bibfnamefont{K.}~\bibnamefont{Freese}}
  (\bibinfo{year}{2008}), \eprint{0808.3607}.

\end{thebibliography}

\end{document}